\documentclass[reprint, superscriptaddress, amsmath, amssymb, aps, prb]{revtex4-2}
\usepackage{graphicx}
\usepackage{dcolumn}
\usepackage{bm}
\usepackage{placeins}
\usepackage{appendix}

\usepackage{verbatim}
\usepackage[usenames,dvipsnames]{xcolor}
 \usepackage{amsmath}
 \usepackage{amsfonts}
 \usepackage{amssymb}
 \usepackage[colorlinks=true,citecolor=blue,linkcolor=red]{hyperref}

 \usepackage{bbold}     
 \usepackage[makeroom]{cancel}  
 \usepackage{multirow}    
 \usepackage[normalem]{ulem}        
 \usepackage{array}
 \usepackage{pdfpages}
\makeatletter
 \AtBeginDocument{\let\LS@rot\@undefined}
 \makeatother

\begin{document}

\title{Large Andreev Bound State Zero Bias Peaks in a Weakly Dissipative Environment}

\author{Zhichuan Wang}
 \email{equal contribution}
\affiliation{Beijing National Laboratory for Condensed Matter Physics, Institute of Physics, Chinese Academy of Sciences, Beijing 100190, China}

\affiliation{State Key Laboratory of Low Dimensional Quantum Physics, Department of Physics, Tsinghua University, Beijing 100084, China}

\affiliation{School of Physical Sciences, University of Chinese Academy of Sciences, Beijing 100049, China}

\author{Shan Zhang}
 \email{equal contribution}
\affiliation{State Key Laboratory of Low Dimensional Quantum Physics, Department of Physics, Tsinghua University, Beijing 100084, China}

\author{Dong Pan}
 \email{equal contribution}
\affiliation{State Key Laboratory of Superlattices and Microstructures, Institute of Semiconductors, Chinese Academy of Sciences, P. O. Box 912, Beijing 100083, China}

\author{Gu Zhang}
 \email{equal contribution}
\affiliation{Beijing Academy of Quantum Information Sciences, 100193 Beijing, China}

\author{Zezhou Xia}
\affiliation{State Key Laboratory of Low Dimensional Quantum Physics, Department of Physics, Tsinghua University, Beijing 100084, China}

\author{Zonglin Li}
\affiliation{State Key Laboratory of Low Dimensional Quantum Physics, Department of Physics, Tsinghua University, Beijing 100084, China}

\author{Donghao Liu}
\affiliation{State Key Laboratory of Low Dimensional Quantum Physics, Department of Physics, Tsinghua University, Beijing 100084, China}

\author{Zhan Cao}
\affiliation{Beijing Academy of Quantum Information Sciences, 100193 Beijing, China}

\author{Lei Liu}
\affiliation{State Key Laboratory of Superlattices and Microstructures, Institute of Semiconductors, Chinese Academy of Sciences, P. O. Box 912, Beijing 100083, China}

\author{Lianjun Wen}
\affiliation{State Key Laboratory of Superlattices and Microstructures, Institute of Semiconductors, Chinese Academy of Sciences, P. O. Box 912, Beijing 100083, China}

\author{Dunyuan Liao}
\affiliation{State Key Laboratory of Superlattices and Microstructures, Institute of Semiconductors, Chinese Academy of Sciences, P. O. Box 912, Beijing 100083, China}

\author{Ran Zhuo}
\affiliation{State Key Laboratory of Superlattices and Microstructures, Institute of Semiconductors, Chinese Academy of Sciences, P. O. Box 912, Beijing 100083, China}

\author{Yongqing Li}
\affiliation{Beijing National Laboratory for Condensed Matter Physics, Institute of Physics, Chinese Academy of Sciences, Beijing 100190, China}
\affiliation{School of Physical Sciences, University of Chinese Academy of Sciences, Beijing 100049, China}

\author{Dong E. Liu}
\affiliation{State Key Laboratory of Low Dimensional Quantum Physics, Department of Physics, Tsinghua University, Beijing 100084, China}
\affiliation{Beijing Academy of Quantum Information Sciences, 100193 Beijing, China}
\affiliation{Frontier Science Center for Quantum Information, 100084 Beijing, China}

\author{Runan Shang}
\email{shangrn@baqis.ac.cn}
\affiliation{Beijing Academy of Quantum Information Sciences, 100193 Beijing, China}

\author{Jianhua Zhao}
 \email{jhzhao@semi.ac.cn}
\affiliation{State Key Laboratory of Superlattices and Microstructures, Institute of Semiconductors, Chinese Academy of Sciences, P. O. Box 912, Beijing 100083, China}

\author{Hao Zhang}
\email{hzquantum@mail.tsinghua.edu.cn}
\affiliation{State Key Laboratory of Low Dimensional Quantum Physics, Department of Physics, Tsinghua University, Beijing 100084, China}
\affiliation{Beijing Academy of Quantum Information Sciences, 100193 Beijing, China}
\affiliation{Frontier Science Center for Quantum Information, 100084 Beijing, China}


\begin{abstract}

We study Andreev bound states in hybrid InAs-Al nanowire devices. The energy of these states can be tuned to zero by gate voltage or magnetic field, revealing large zero bias peaks (ZBPs) near $2e^2/h$ in tunneling conductance. Probing these large ZBPs using a weakly dissipative lead reveals non-Fermi liquid temperature ($T$) dependence due to environmental Coulomb blockade (ECB), an interaction effect from the lead acting on the nanowire junction. By increasing $T$, these large ZBPs either show a height increase or a transition from split peaks to a ZBP, both deviate significantly from non-dissipative devices where a Fermi-liquid $T$ dependence is revealed. Our result demonstrates the competing effect between ECB and thermal broadening on Andreev bound states.

\end{abstract}

\maketitle

Andreev bound states (ABSs) can emerge in non-uniform superconductors by Andreev scattering at energies below the superconducting gap \cite{Andreev1965, Andreev1966}. Hybrid semiconductor-superconductor systems \cite{Pillet2010, Mason2011, Chang2013, Jiangyuying} provide an ideal test-bed to study these subgap states, thanks to the proximity effect mediated by Andreev reflections and the high tunability of carrier density using electrostatic gates. Fascinating physics can be revealed by adding additional elements, e.g. one dimensionality and spin-oribt coupling \cite{vanDam2006, Silvano2014, Zhang2017Ballistic, Michiel2018, Jouri2019, Schonenberger2020}. These hybrid nanowires, with the semiconductor being InAs or InSb, are further predicted to host Majorana zero modes (MZMs) \cite{Lutchyn2010, Oreg2010} where one ABS can be spatially separated into two `halves' (MZMs). In tunneling conductance, zero bias peaks (ZBPs) can be observed \cite{Mourik, Silvano2014, Deng2016, Gul2018, Zhang2021, Song2021, Prada2020} as a possible signature for MZMs as well as zero-energy ABSs. The similarities between ABSs and MZMs create huge debates on distinguishing them \cite{Prada2012, Patrick_Lee_disorder_2012, BrouwerSmooth, Liu2017, TudorQuasi, WimmerQuasi, DEL-Disorder2018, CaoZhanPRL,Loss2018ABS, GoodBadUgly, DasSarma2021Disorder,Tudor2021Disorder,2021_PRB_Donghao}.

Recently, motivated by a theoretical work \cite{Dong_PRL2013}, we have added a resistive lead as a strongly dissipative environment to the hybrid InAs-Al devices \cite{ZhangShan}. Previously, dissipative tunneling of metallic junctions \cite{Delsing_1989, Ingold, Flensberg_1992, Joyez_1998, Zheng_1998} and semiconductor nanostructures \cite{Pierre2011,Gleb_Nature, Gleb_NaturePhysics,Dong_PRB2014, Jezouin_2013,Pierre_PRX} have been widely studied, exhibiting environmental Coulomb blockade (ECB) with power laws emulating Luttinger liquid physics \cite{Safi_2004}. Here \cite{ZhangShan}, the interaction effect in the environment acts on the InAs-Al nanowire junction where ABSs emerge. In the strongly dissipative regime, we have shown that most zero-energy ABSs are revealed as split peaks instead of ZBPs \cite{ZhangShan}. However, it still remains as an interesting question to ask how do ABS-induced ZBPs, if not being fully suppressed, behave in a dissipative environment. In this work, we lower the dissipation strength by reducing the lead resistance from $\sim$ 5.7 k$\Omega$ in Ref. \cite{ZhangShan} to $\sim$ 2.7 k$\Omega$. In this weak dissipation regime, we can resolve large (trivial) ZBPs near $2e^2/h$, the main focus of the current debates \cite{GoodBadUgly, DasSarma2021Disorder,Tudor2021Disorder}. $T$ dependence of these large ZBPs behave dramatically different from that of small ZBPs or ZBPs in regular devices without dissipation. We ascribe this difference to the dynamical competition between ECB and thermal averaging. Our result sets a lower bar on dissipation strength if being used to distinguish MZMs \cite{Dong_PRL2013}.

Fig. 1a shows the scanning electron micrograph (SEM) of Device A. An InAs nanowire (gray) with a thin Al shell (cyan) is first contacted by Ti/Au (yellow), and then connected to a resistive film (red), serving as the dissipative environment. Resistance of the dissipative film is designed and later estimated to be $\sim$ 2.7 k$\Omega$ (see Fig. S1 for details). The device can be tuned by a side tunnel gate (TG) and a global back gate (BG) which is p-doped Si covered by 300 nm thick SiO$_2$. Growth and transport details of these hybrid InAs-Al nanowires can be found in Ref. \cite{Pan2020, Song2021}. We apply a total bias voltage ($V_{\text{bias}}$) on the left Ti/Au lead, and measure the current $I$ from the right contact. The bias drop on the InAs-Al part is $V=V_{\text{bias}}-I\times R_{\text{series}}$, where $R_{\text{series}}=R_{\text{filters}}+R_{\text{film}}$ includes resistance of the fridge filters and the dissipative film, both estimated based on independent calibration. The device differential conductance $G \equiv dI/dV=(dV_{\text{bias}}/dI-R_{\text{series}})^{-1}$ has $R_{\text{series}}$ excluded. 

\begin{figure*}[htb]
\includegraphics[width=\textwidth]{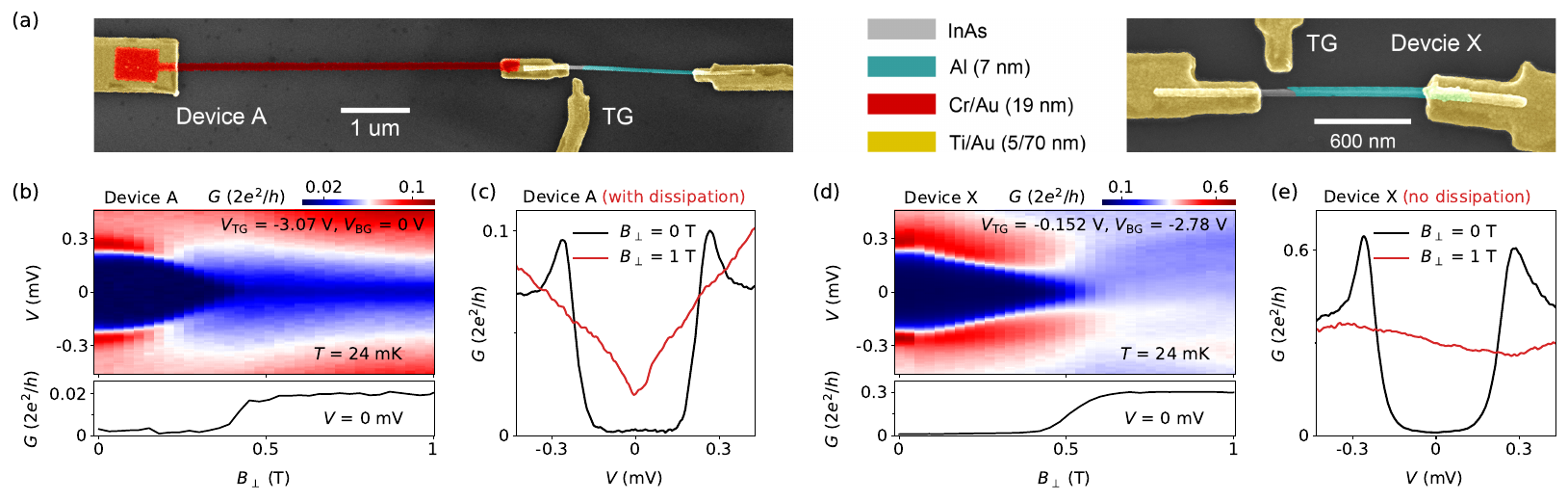}
\centering
\caption{(a) False color SEM of Device A (left) and control Device X (right). See labeling for Al, Cr/Au and Ti/Au film thickness. Device X is the same device used in Ref \cite{ZhangShan}, therefore the SEM is re-used here. (b) $B$ scan of the superconducting gap in Device A. $B$ is in-plane and perpendicular to the wire axis. Lower, zero-bias line cut. (c) Line cuts from (b) at 0 T and 1 T. (d-e) $B$ scan of the gap in Device X. $T\sim$ 24 mK.}
\label{fig1}
\end{figure*}

In Fig. 1b we tune the device into tunneling regime and resolve the superconducting gap (see Fig. S2 for a full gate scan). The gap is closed by a magnetic field ($B$), perpendicular to the nanowire, at $\sim$ 0.5 T. After the gap closing, a dip is resolved near zero-bias (red curve in Fig. 1c). This suppression of $G$ near zero-bias (see Fig. S3 more scans) is a signature of ECB. We further quantify this suppression with power laws and find rough matches (see Fig. S4). The small deviations from power law fitting indicate an incomplete ECB suppression of imperfections, e.g. defects and local states in the InAs-Al junction. In our control Device X, an InAs-Al nanowire without the dissipative resistor, $G$ is usually flat after the gap closing without noticeable features near zero-bias as shown in Fig. 1de. 

\begin{figure*}[tb]
\includegraphics[width=\textwidth]{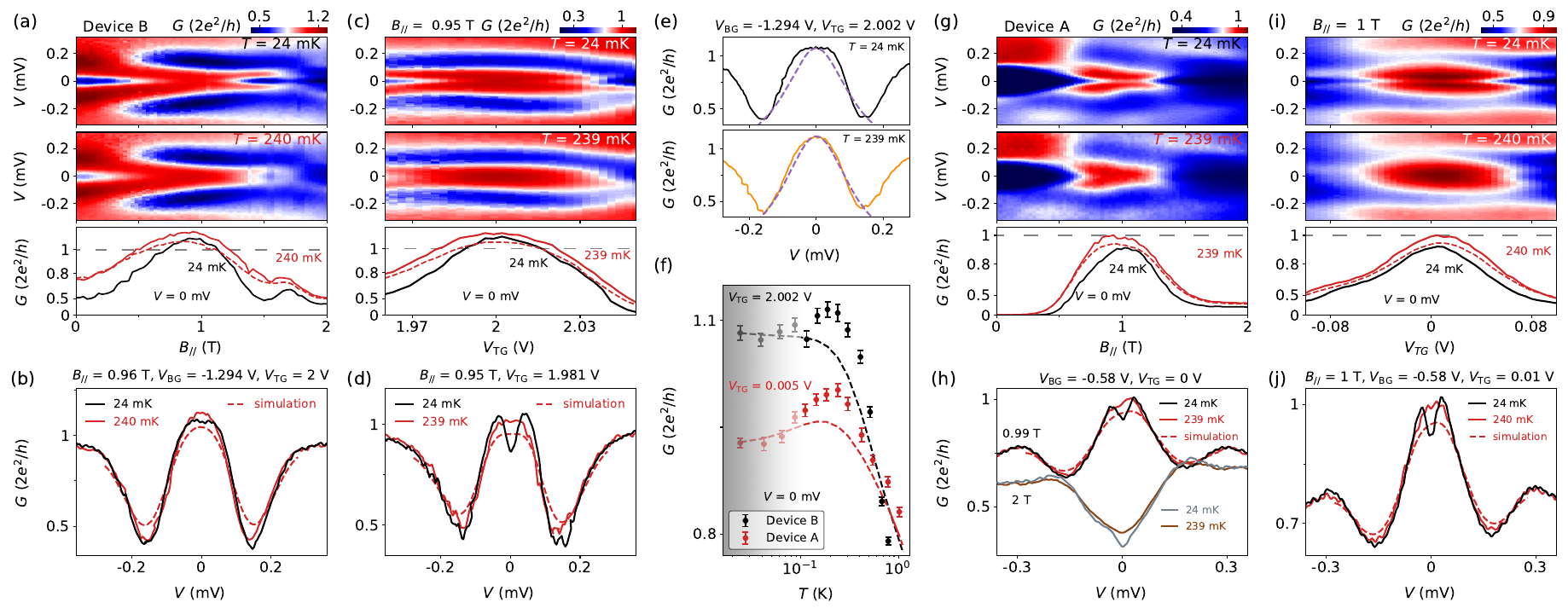}
\centering
\caption{(a) $G$ vs $V$ and $B$ at $T$ = 24 mK (upper) and 240 mK (middle) for Device B. Lower panel: zero-bias line cuts, and thermal simulation for $T$ = 240 mK (red dashed line). (b) Line cuts from (a) at $B$ = 0.96 T and thermal simulation (dashed line) for 240 mK. (c) $V_{\text{TG}}$ scan at $B=0.95$ T for 24 mK (upper) and 239 mK (middle). Lower-panel, zero-bias line cuts and thermal simulation (dashed line). (d) Line cuts from (c) and thermal simulation (dashed line) for 239 mK. (e) A ZBP line cut from (c) at 24 mK  (upper) and 239 mK (lower). Dashed lines are the Lorentzian fits assuming 24 mK and 239 mK thermal broadening. (f) $T$ dependence of zero-bias $G$. The dashed line is the thermal simulation. Black for Device B and red for Device A from (g-j). (g-j) Similar to (a-d) but for Device A. In the lower panels of (a), (c), (g) and (i), the y-axis is scaled linear for $G^2$ instead of $G$, to better resolve the deviations at higher $G$s. }
\label{fig2}
\end{figure*}

We now align $B$ parallel to the nanowire and find large ZBPs due to zero-energy ABSs. Fig. 2a shows a $B$ scan of an ABS, likely disorder-induced, in Device B at temperatures ($T$) of 24 mK and 240 mK. $T$ refers to the fridge $T$ unless specified. Device B has a dissipative resistor ($\sim$ 2.7 k$\Omega$) similar to that in Device A. At 0.96 T, two peaks merge at zero and form a large ZBP with its height exceeding $2e^2/h$, see Fig. 2b line cuts. Interestingly, the ZBP height at an elevated $T$ of 240 mK (red curve) is higher than that at base $T$ of 24 mK (black). This unusual $T$ dependence is qualitatively different from those in non-dissipative devices where the ZBP height always decreases as $T$ increases, a purely thermal broadening (averaging) effect (see Fig. S5 examples in control Device X). We attribute this unusual $T$ dependence to the mixing or competing effect of ECB and thermal broadening. ECB suppresses zero-bias $G$ (the ZBP height) at low $T$ (24 mK). Higher $T$ (240 mK) diminishes ECB and enhances the height. On the contrary, thermal broadening, an averaging effect over a peak, leads to the opposite trend in $T$ dependence. For large and broad ZBPs, thermal broadening effect is not obvious at low $T$, therefore ECB may dominate the trend in $T$ dependence, as visualized by the difference between the black and the red curves in Fig. 2b. Note that ECB is the strongest near zero $V$.

To simulate the thermal broadening effect, we use the formula $G(V,T)=\int_{-\infty}^{+\infty}G(\epsilon,0) \frac{\partial f(eV-\epsilon, T)}{\partial \epsilon} d\epsilon$, where $f(E,T)=\frac{1}{e^{E/k_BT}+1}$ is the Fermi distribution function. $G(V,T)$ at high $T$ can be calculated by this convolution using $G(V,T$ = 0 K$)$ as an input which we replace with $G(V, T$ = 24 mK$)$. This assumption should be valid for $T$ much larger than 24 mK. The red dashed lines in Fig. 2a (lower panel) and Fig. 2b are the simulation results for $T$ = 240 mK, noticeably lower than the measured $G$ at the same $T$ (red lines). This deviation suggests that thermal broadening is not the only effect and ECB should be included in the $T$ dependence of large ZBPs.

Fig. 2c shows the gate dependence of the large ZBP where again a sizable deviation can be found between the measurement (red line) and the thermal simulation (dashed line) for $T$ = 239 mK. Fig. 2d shows a line cut of the near-zero-energy ABS: split peaks at 24 mK evolving into a large ZBP at 239 mK. The red dashed line is the thermal simulation which could also merge split peaks into a ZBP but at a cost of lowering the peak height.

ECB not only suppresses the ZBP height, but also modifies the peak shape due to its non-uniform suppression over bias. In Fig. 2e, the large ZBP at base $T$ shows larger deviation from the Lorentzian fit (dashed line) than that at higher $T$. This phenomenon is expected since ECB is stronger (weaker) at lower (higher) $T$. Fig. 2f shows the full $T$ dependence of the zero-bias $G$ (black dots) for this ZBP. The black dashed line is the thermal simulation which, as expected, shows a monotonic decrease with increasing $T$. Contrarily, the measured ZBP height first increases (ECB being weakened) and then decreases until $T$ being too high where thermal broadening starts dominating. We sketch a gradient gray background for $T<$ 100 mK indicating that below which the electron $T$ gradually deviates from the fridge $T$ and finally saturates. 

\begin{figure}[tb]
\includegraphics[width=\columnwidth]{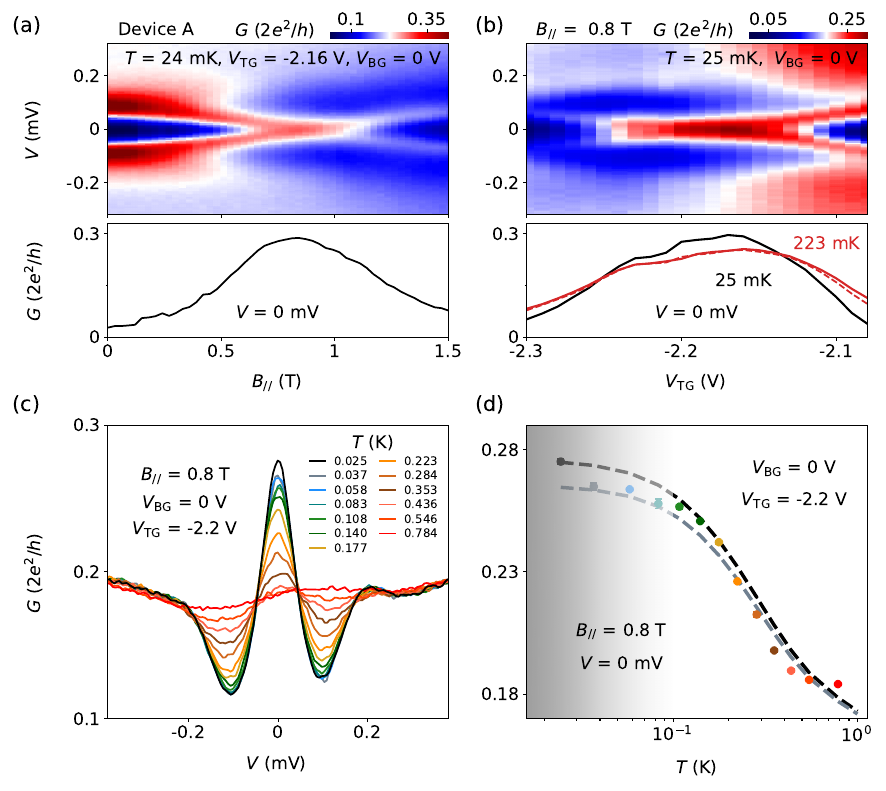}
\centering
\caption{(a) $B$ dependence of a small ZBP in Device A at 24 mK. Lower panel, zero-bias line cut. (b) $V_{\text{TG}}$ scan of this ZBP at $B$ = 0.8 T. $T$ = 25 mK. Lower panel, zero-bias line cuts at 25 mK (black) and 223 mK (red curve), together with thermal simulation (red dashed line). (c) $T$ dependence of the ZBP. (d) Zero-bias $G$ (dots) from (c). The black and gray dashed lines are thermal simulations using the 25 mK and 37 mK data as the input for $G(V, 0)$, respectively, to account for possible instabilities during measurement. }
\label{fig3}
\end{figure}

Similar ZBPs near $2e^2/h$ (with a smaller above-background-peak-height) can also be observed in Device A, as shown in Fig. 2g-j. Differently, this zero-energy ABS resolves a small splitting at base $T$ (black curves in Fig. 2hj). This splitting feature has no fundamental difference from the large (non-split) ZBP in Fig. 2b (black curve). Splitting or not depends on the zero-energy ABS details, dissipation strength and $T$. If the electron $T$ in Fig. 2b could be further lowered towards zero, ECB would suppress more and the ZBP in Fig. 2b would also split. If the dissipation strength was increased towards the strong dissipation regime, most zero-energy ABS induced ZBPs would split. In fact, in Fig. 2a (for $B$ slightly different from 0.96 T) and Fig. 2d, we could also find large ZBPs at higher $T$s which resolve a small splitting at base $T$. Nevertheless, compared with Fig. 2b, Fig. 2h demonstrates a different regime for zero-energy ABS with a small splitting at base $T$. The main noticeable difference between base $T$ (black curve) and higher $T$ (red curve) in Fig. 2hj is the `triangle' area near zero-bias, a visualization of ECB (also visible for the 2 T line cuts). Note that thermal simulation (red dashed lines) conserves the area underneath the curve, making the peak height lower and peak width wider for higher $T$s.

The red dots in Fig. 2f illustrate the $T$ evolution of the zero-bias $G$ corresponding to this zero-energy ABS. With increasing $T$, the zero-bias $G$ first increases and then decreases, similar to the black dots. Though if using a split-peak at base $T$ as the input, thermal simulation would also give an initial increase of $G$ (purely due to averaging effect on a dip) as shown by the red dashed line. This increase is much less than the measurement data, indicating the noticeable role of ECB. For more $T$ dependence of the ABSs in Fig. 2, see Fig. S6.

Above we have demonstrated the interplay between weak dissipation (ECB) and thermal broadening on large ZBPs induced by zero-energy ABSs. These large ZBPs generally have a large peak width, therefore immune to thermal broadening for $T$s being not too high (e.g. below 300 mK). Within this range, increasing $T$ diminishes ECB and enhances the zero-bias $G$, causing deviations from the Fermi-liquid $T$ dependence (thermal simulation).

In Fig. 3 we study small ZBPs in Device A under the same dissipation strength. Fig. 3ab show the $B$ and gate scan of an ABS at base $T$. The level crossing point, corresponding to a zero-energy ABS, resolves a small ZBP (peak height $\sim 0.3\times 2e^2/h$). The ZBP width is also narrower than those in Fig. 2, therefore more sensitive to thermal averaging. Indeed, in Fig. 3b (lower panel), the thermal simulation (dashed line) for $T$ = 223 mK agrees reasonably well with the measurement (red line), suggesting that thermal broadening is the dominating effect, different from the large ZBP case in Fig. 2.

Fig. 3c and 3d show the full $T$ evolution of this small ZBP. The zero-bias $G$ (dots in Fig. 3d) shows a monotonic decrease as increasing $T$, qualitatively different from the trend of $T$ dependence in Fig. 2. Moreover, thermal simulation (both black and gray dashed lines in Fig. 3d) matches reasonable well with the measurement, confirming thermal averaging as the dominating effect for small ZBPs. Note that weak dissipation should still be present for this ABS since they all share the same dissipative resistor. The narrower peak widths for small ZBPs make them more sensitive to thermal averaging. This causes ECB effect almost unnoticeable in the $T$ evolution. For more $T$ dependence and gate scans of this ABS, see Fig. S7. In Fig. S8, we present some stability tests of the ABSs in Fig. 2 and Fig. 3 to rule out possible gate shifts or charge jumps. Fig. S9-S11 show four more zero-energy ABSs with intermediate ZBP heights to illustrate the gradual transition from large to small ZBPs. 

\begin{figure}[t]
\includegraphics[width=0.9\columnwidth]{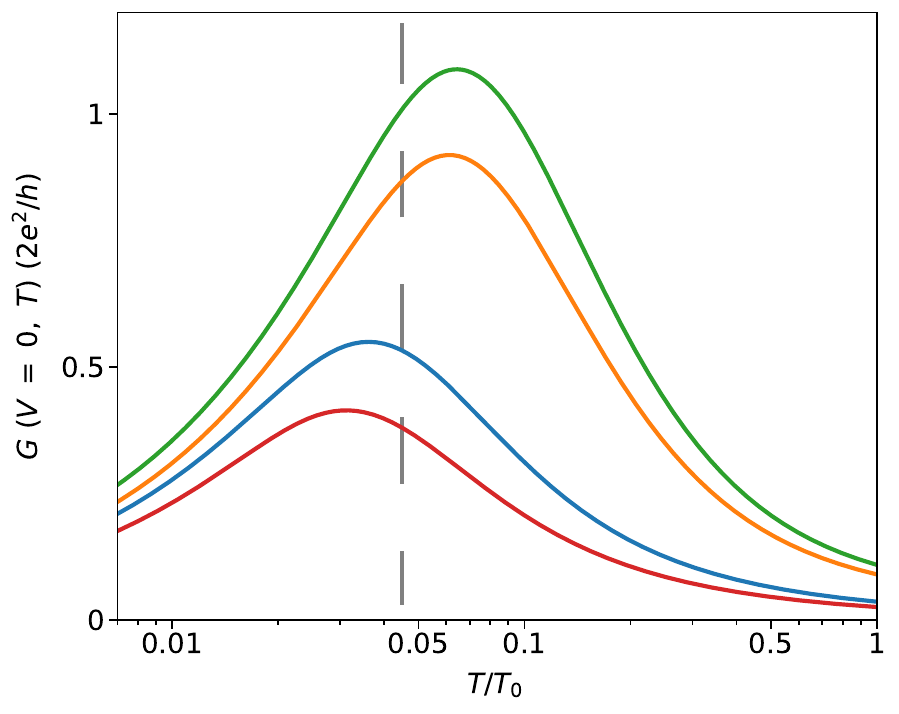}
\centering
\caption{Renormalization group (RG) calculation of $G$ for various zero-energy ABSs (different colors), corresponding to different ZBP heights. The dissipation strength $r$ = 0.1 for all curves. $T_0$ refers to the starting RG temperature. We assign the base $T$ in experiment to the vertical dashed line. }
\label{fig4}
\end{figure}

In Fig. 4, we present the result of renormalization group (RG) calculation for zero-energy ABSs by fixing the dissipation strength $r$ = 0.1. This $r$ translates to an effective dissipative resistance of $r\times h/e^2\sim$ 2.58 k$\Omega$, similar to the case of Device A and B. In the theory model \cite{Dong_2021}, we could modify the lead-ABS tunnel couplings for electrons ($t_\text{e}$) and Andreev reflected holes ($t_\text{h}$) to realize different ZBP heights: if $t_\text{e}$ and $t_\text{h}$ are larger and closer, the ZBP height is larger. Fig. 4 demonstrates the $T$ dependence of zero-energy ABSs with different heights (colors). When $T$ decreases towards absolute zero, the zero-bias $G$ should be suppressed towards zero due to dissipation for all ABS cases, causing ZBP splitting. When $T$ is high enough and increases, ECB is diminished and the ZBP is thermally smeared, also causing the zero-bias $G$ to decrease. These two regions combined together form the shape of $T$ dependence in Fig. 4 with a maximum peak at an intermediate $T$. The key point is that this intermediate $T$ is different for large and small ZBPs. By comparing with experimental data, we think our base $T$ likely corresponds the vertical dashed line. Under this assumption, for small ZBPs (the red and blue curves), the zero-bias $G$ shows a continuous decrease as increasing $T$, consistent with the observation in Fig. 3. For large ZBPs (the orange and green curves), the zero-bias $G$ initially increases and then decreases as increasing $T$, consistent with the observation in Fig. 2. 

To summarize, we have studied zero-energy ABSs in a weakly dissipative environment. Large ZBPs near $2e^2/h$ can be resolved. The large peak width protects the ZBP height from decreasing against thermal averaging over an intermediate $T$-range (e.g. $<$ 300 mK) where ECB could dominate. As a result, when $T$ is lowered, the ZBP could either decrease its height or split, both not following Fermi-liquid thermal simulation. On the contrary, $T$ dependence of small ZBPs follows thermal simulation, same with the cases in regular devices without dissipation. Our result shows that weak dissipation strength ($R_{\text{film}}\sim$ 2.7 k$\Omega$) can not suppress ABS-induced ZBPs, therefore sets a lower bar on dissipation strength for future MZM devices \cite{NextSteps}. For Majorana or quasi-Majorana resonance detection \cite{TudorQuasi, WimmerQuasi}, stronger dissipation regime \cite{ZhangShan} is preferred which can effectively `filter out' trivial ZBPs \cite{Dong_PRL2013,Gu_2020, Dong_PRB2020, Dong_2021}. Since in the weak dissipation regime, the heights of ZBPs induced by zero-energy ABSs could either increase (Fig. 2) or decrease (Fig. 3) as increasing $T$, depending on the ABS details. This makes it difficult to distinguish MZM signatures \cite{DasSarma2001, Law2009} solely based on the $T$ dependence \cite{Dong_PRL2013}.

\textbf{Acknowledgment} Raw data and processing codes within this paper are available at https://doi.org/10.5281/zenodo.6123849. This work is supported by Tsinghua University Initiative Scientific Research Program, Alibaba Innovative Research Program, National Natural Science Foundation of China (Grant No. 92065106, 61974138, 11974198, and 12004040), Beijing Natural Science Foundation (Grant No. 1192017). D. P. acknowledges the support from Youth Innovation Promotion Association, Chinese Academy of Sciences (No. 2017156).

\bibliography{mybibfile}

\begin{thebibliography}{54}%
\makeatletter
\providecommand \@ifxundefined [1]{%
 \@ifx{#1\undefined}
}%
\providecommand \@ifnum [1]{%
 \ifnum #1\expandafter \@firstoftwo
 \else \expandafter \@secondoftwo
 \fi
}%
\providecommand \@ifx [1]{%
 \ifx #1\expandafter \@firstoftwo
 \else \expandafter \@secondoftwo
 \fi
}%
\providecommand \natexlab [1]{#1}%
\providecommand \enquote  [1]{``#1''}%
\providecommand \bibnamefont  [1]{#1}%
\providecommand \bibfnamefont [1]{#1}%
\providecommand \citenamefont [1]{#1}%
\providecommand \href@noop [0]{\@secondoftwo}%
\providecommand \href [0]{\begingroup \@sanitize@url \@href}%
\providecommand \@href[1]{\@@startlink{#1}\@@href}%
\providecommand \@@href[1]{\endgroup#1\@@endlink}%
\providecommand \@sanitize@url [0]{\catcode `\\12\catcode `\$12\catcode
  `\&12\catcode `\#12\catcode `\^12\catcode `\_12\catcode `\%12\relax}%
\providecommand \@@startlink[1]{}%
\providecommand \@@endlink[0]{}%
\providecommand \url  [0]{\begingroup\@sanitize@url \@url }%
\providecommand \@url [1]{\endgroup\@href {#1}{\urlprefix }}%
\providecommand \urlprefix  [0]{URL }%
\providecommand \Eprint [0]{\href }%
\providecommand \doibase [0]{https://doi.org/}%
\providecommand \selectlanguage [0]{\@gobble}%
\providecommand \bibinfo  [0]{\@secondoftwo}%
\providecommand \bibfield  [0]{\@secondoftwo}%
\providecommand \translation [1]{[#1]}%
\providecommand \BibitemOpen [0]{}%
\providecommand \bibitemStop [0]{}%
\providecommand \bibitemNoStop [0]{.\EOS\space}%
\providecommand \EOS [0]{\spacefactor3000\relax}%
\providecommand \BibitemShut  [1]{\csname bibitem#1\endcsname}%
\let\auto@bib@innerbib\@empty
\bibitem [{\citenamefont {Andreev}(1965)}]{Andreev1965}%
  \BibitemOpen
  \bibfield  {author} {\bibinfo {author} {\bibfnamefont {A.}~\bibnamefont
  {Andreev}},\ }\bibfield  {title} {\bibinfo {title} {Thermal conductivity of
  the intermediate state of superconductors ii},\ }\href@noop {} {\bibfield
  {journal} {\bibinfo  {journal} {Sov. Phys. JETP}\ }\textbf {\bibinfo {volume}
  {20}},\ \bibinfo {pages} {1490} (\bibinfo {year} {1965})}\BibitemShut
  {NoStop}%
\bibitem [{\citenamefont {Andreev}(1966)}]{Andreev1966}%
  \BibitemOpen
  \bibfield  {author} {\bibinfo {author} {\bibfnamefont {A.}~\bibnamefont
  {Andreev}},\ }\bibfield  {title} {\bibinfo {title} {Electron spectrum of the
  intermediate state of superconductors},\ }\href@noop {} {\bibfield  {journal}
  {\bibinfo  {journal} {Sov. Phys. JETP}\ }\textbf {\bibinfo {volume} {22}},\
  \bibinfo {pages} {18} (\bibinfo {year} {1966})}\BibitemShut {NoStop}%
\bibitem [{\citenamefont {Pillet}\ \emph {et~al.}(2010)\citenamefont {Pillet},
  \citenamefont {Quay}, \citenamefont {Morfin}, \citenamefont {Bena},
  \citenamefont {Yeyati},\ and\ \citenamefont {Joyez}}]{Pillet2010}%
  \BibitemOpen
  \bibfield  {author} {\bibinfo {author} {\bibfnamefont {J.}~\bibnamefont
  {Pillet}}, \bibinfo {author} {\bibfnamefont {C.}~\bibnamefont {Quay}},
  \bibinfo {author} {\bibfnamefont {P.}~\bibnamefont {Morfin}}, \bibinfo
  {author} {\bibfnamefont {C.}~\bibnamefont {Bena}}, \bibinfo {author}
  {\bibfnamefont {A.~L.}\ \bibnamefont {Yeyati}},\ and\ \bibinfo {author}
  {\bibfnamefont {P.}~\bibnamefont {Joyez}},\ }\bibfield  {title} {\bibinfo
  {title} {Andreev bound states in supercurrent-carrying carbon nanotubes
  revealed},\ }\href@noop {} {\bibfield  {journal} {\bibinfo  {journal} {Nature
  Physics}\ }\textbf {\bibinfo {volume} {6}},\ \bibinfo {pages} {965} (\bibinfo
  {year} {2010})}\BibitemShut {NoStop}%
\bibitem [{\citenamefont {Dirks}\ \emph {et~al.}(2011)\citenamefont {Dirks},
  \citenamefont {Hughes}, \citenamefont {Lal}, \citenamefont {Uchoa},
  \citenamefont {Chen}, \citenamefont {Chialvo}, \citenamefont {Goldbart},\
  and\ \citenamefont {Mason}}]{Mason2011}%
  \BibitemOpen
  \bibfield  {author} {\bibinfo {author} {\bibfnamefont {T.}~\bibnamefont
  {Dirks}}, \bibinfo {author} {\bibfnamefont {T.~L.}\ \bibnamefont {Hughes}},
  \bibinfo {author} {\bibfnamefont {S.}~\bibnamefont {Lal}}, \bibinfo {author}
  {\bibfnamefont {B.}~\bibnamefont {Uchoa}}, \bibinfo {author} {\bibfnamefont
  {Y.-F.}\ \bibnamefont {Chen}}, \bibinfo {author} {\bibfnamefont
  {C.}~\bibnamefont {Chialvo}}, \bibinfo {author} {\bibfnamefont {P.~M.}\
  \bibnamefont {Goldbart}},\ and\ \bibinfo {author} {\bibfnamefont
  {N.}~\bibnamefont {Mason}},\ }\bibfield  {title} {\bibinfo {title} {Transport
  through andreev bound states in a graphene quantum dot},\ }\href@noop {}
  {\bibfield  {journal} {\bibinfo  {journal} {Nature Physics}\ }\textbf
  {\bibinfo {volume} {7}},\ \bibinfo {pages} {386} (\bibinfo {year}
  {2011})}\BibitemShut {NoStop}%
\bibitem [{\citenamefont {Chang}\ \emph {et~al.}(2013)\citenamefont {Chang},
  \citenamefont {Manucharyan}, \citenamefont {Jespersen}, \citenamefont
  {Nyg{\aa}rd},\ and\ \citenamefont {Marcus}}]{Chang2013}%
  \BibitemOpen
  \bibfield  {author} {\bibinfo {author} {\bibfnamefont {W.}~\bibnamefont
  {Chang}}, \bibinfo {author} {\bibfnamefont {V.}~\bibnamefont {Manucharyan}},
  \bibinfo {author} {\bibfnamefont {T.}~\bibnamefont {Jespersen}}, \bibinfo
  {author} {\bibfnamefont {J.}~\bibnamefont {Nyg{\aa}rd}},\ and\ \bibinfo
  {author} {\bibfnamefont {C.}~\bibnamefont {Marcus}},\ }\bibfield  {title}
  {\bibinfo {title} {Tunneling spectroscopy of quasiparticle bound states in a
  spinful josephson junction},\ }\href@noop {} {\bibfield  {journal} {\bibinfo
  {journal} {Physical Review Letters}\ }\textbf {\bibinfo {volume} {110}},\
  \bibinfo {pages} {217005} (\bibinfo {year} {2013})}\BibitemShut {NoStop}%
\bibitem [{\citenamefont {Jiang}\ \emph {et~al.}(2021)\citenamefont {Jiang},
  \citenamefont {Yang}, \citenamefont {Li}, \citenamefont {Song}, \citenamefont
  {Miao}, \citenamefont {Tong}, \citenamefont {Geng}, \citenamefont {Gao},
  \citenamefont {Li}, \citenamefont {Zhang} \emph {et~al.}}]{Jiangyuying}%
  \BibitemOpen
  \bibfield  {author} {\bibinfo {author} {\bibfnamefont {Y.}~\bibnamefont
  {Jiang}}, \bibinfo {author} {\bibfnamefont {S.}~\bibnamefont {Yang}},
  \bibinfo {author} {\bibfnamefont {L.}~\bibnamefont {Li}}, \bibinfo {author}
  {\bibfnamefont {W.}~\bibnamefont {Song}}, \bibinfo {author} {\bibfnamefont
  {W.}~\bibnamefont {Miao}}, \bibinfo {author} {\bibfnamefont {B.}~\bibnamefont
  {Tong}}, \bibinfo {author} {\bibfnamefont {Z.}~\bibnamefont {Geng}}, \bibinfo
  {author} {\bibfnamefont {Y.}~\bibnamefont {Gao}}, \bibinfo {author}
  {\bibfnamefont {R.}~\bibnamefont {Li}}, \bibinfo {author} {\bibfnamefont
  {Q.}~\bibnamefont {Zhang}}, \emph {et~al.},\ }\bibfield  {title} {\bibinfo
  {title} {Selective area epitaxy of pbte-pb hybrid nanowires on a
  lattice-matched substrate},\ }\href@noop {} {\bibfield  {journal} {\bibinfo
  {journal} {arXiv preprint arXiv:2110.13642}\ } (\bibinfo {year}
  {2021})}\BibitemShut {NoStop}%
\bibitem [{\citenamefont {van Dam}\ \emph {et~al.}(2006)\citenamefont {van
  Dam}, \citenamefont {Nazarov}, \citenamefont {Bakkers}, \citenamefont
  {De~Franceschi},\ and\ \citenamefont {Kouwenhoven}}]{vanDam2006}%
  \BibitemOpen
  \bibfield  {author} {\bibinfo {author} {\bibfnamefont {J.~A.}\ \bibnamefont
  {van Dam}}, \bibinfo {author} {\bibfnamefont {Y.~V.}\ \bibnamefont
  {Nazarov}}, \bibinfo {author} {\bibfnamefont {E.~P.}\ \bibnamefont
  {Bakkers}}, \bibinfo {author} {\bibfnamefont {S.}~\bibnamefont
  {De~Franceschi}},\ and\ \bibinfo {author} {\bibfnamefont {L.~P.}\
  \bibnamefont {Kouwenhoven}},\ }\bibfield  {title} {\bibinfo {title}
  {Supercurrent reversal in quantum dots},\ }\href@noop {} {\bibfield
  {journal} {\bibinfo  {journal} {Nature}\ }\textbf {\bibinfo {volume} {442}},\
  \bibinfo {pages} {667} (\bibinfo {year} {2006})}\BibitemShut {NoStop}%
\bibitem [{\citenamefont {Lee}\ \emph {et~al.}(2014)\citenamefont {Lee},
  \citenamefont {Jiang}, \citenamefont {Houzet}, \citenamefont {Aguado},
  \citenamefont {Lieber},\ and\ \citenamefont {De~Franceschi}}]{Silvano2014}%
  \BibitemOpen
  \bibfield  {author} {\bibinfo {author} {\bibfnamefont {E.~J.}\ \bibnamefont
  {Lee}}, \bibinfo {author} {\bibfnamefont {X.}~\bibnamefont {Jiang}}, \bibinfo
  {author} {\bibfnamefont {M.}~\bibnamefont {Houzet}}, \bibinfo {author}
  {\bibfnamefont {R.}~\bibnamefont {Aguado}}, \bibinfo {author} {\bibfnamefont
  {C.~M.}\ \bibnamefont {Lieber}},\ and\ \bibinfo {author} {\bibfnamefont
  {S.}~\bibnamefont {De~Franceschi}},\ }\bibfield  {title} {\bibinfo {title}
  {Spin-resolved andreev levels and parity crossings in hybrid
  superconductor--semiconductor nanostructures},\ }\href@noop {} {\bibfield
  {journal} {\bibinfo  {journal} {Nature Nanotechnology}\ }\textbf {\bibinfo
  {volume} {9}},\ \bibinfo {pages} {79} (\bibinfo {year} {2014})}\BibitemShut
  {NoStop}%
\bibitem [{\citenamefont {Zhang}\ \emph {et~al.}(2017)\citenamefont {Zhang},
  \citenamefont {G{\"u}l}, \citenamefont {Conesa-Boj}, \citenamefont {Nowak},
  \citenamefont {Wimmer}, \citenamefont {Zuo}, \citenamefont {Mourik},
  \citenamefont {De~Vries}, \citenamefont {Van~Veen}, \citenamefont {De~Moor}
  \emph {et~al.}}]{Zhang2017Ballistic}%
  \BibitemOpen
  \bibfield  {author} {\bibinfo {author} {\bibfnamefont {H.}~\bibnamefont
  {Zhang}}, \bibinfo {author} {\bibfnamefont {{\"O}.}~\bibnamefont {G{\"u}l}},
  \bibinfo {author} {\bibfnamefont {S.}~\bibnamefont {Conesa-Boj}}, \bibinfo
  {author} {\bibfnamefont {M.~P.}\ \bibnamefont {Nowak}}, \bibinfo {author}
  {\bibfnamefont {M.}~\bibnamefont {Wimmer}}, \bibinfo {author} {\bibfnamefont
  {K.}~\bibnamefont {Zuo}}, \bibinfo {author} {\bibfnamefont {V.}~\bibnamefont
  {Mourik}}, \bibinfo {author} {\bibfnamefont {F.~K.}\ \bibnamefont
  {De~Vries}}, \bibinfo {author} {\bibfnamefont {J.}~\bibnamefont {Van~Veen}},
  \bibinfo {author} {\bibfnamefont {M.~W.}\ \bibnamefont {De~Moor}}, \emph
  {et~al.},\ }\bibfield  {title} {\bibinfo {title} {Ballistic superconductivity
  in semiconductor nanowires},\ }\href@noop {} {\bibfield  {journal} {\bibinfo
  {journal} {Nature Communications}\ }\textbf {\bibinfo {volume} {8}},\
  \bibinfo {pages} {1} (\bibinfo {year} {2017})}\BibitemShut {NoStop}%
\bibitem [{\citenamefont {de~Moor}\ \emph {et~al.}(2018)\citenamefont
  {de~Moor}, \citenamefont {Bommer}, \citenamefont {Xu}, \citenamefont
  {Winkler}, \citenamefont {Antipov}, \citenamefont {Bargerbos}, \citenamefont
  {Wang}, \citenamefont {Van~Loo}, \citenamefont {het Veld}, \citenamefont
  {Gazibegovic} \emph {et~al.}}]{Michiel2018}%
  \BibitemOpen
  \bibfield  {author} {\bibinfo {author} {\bibfnamefont {M.~W.}\ \bibnamefont
  {de~Moor}}, \bibinfo {author} {\bibfnamefont {J.~D.}\ \bibnamefont {Bommer}},
  \bibinfo {author} {\bibfnamefont {D.}~\bibnamefont {Xu}}, \bibinfo {author}
  {\bibfnamefont {G.~W.}\ \bibnamefont {Winkler}}, \bibinfo {author}
  {\bibfnamefont {A.~E.}\ \bibnamefont {Antipov}}, \bibinfo {author}
  {\bibfnamefont {A.}~\bibnamefont {Bargerbos}}, \bibinfo {author}
  {\bibfnamefont {G.}~\bibnamefont {Wang}}, \bibinfo {author} {\bibfnamefont
  {N.}~\bibnamefont {Van~Loo}}, \bibinfo {author} {\bibfnamefont {R.~L.~O.}\
  \bibnamefont {het Veld}}, \bibinfo {author} {\bibfnamefont {S.}~\bibnamefont
  {Gazibegovic}}, \emph {et~al.},\ }\bibfield  {title} {\bibinfo {title}
  {Electric field tunable superconductor-semiconductor coupling in majorana
  nanowires},\ }\href@noop {} {\bibfield  {journal} {\bibinfo  {journal} {New
  Journal of Physics}\ }\textbf {\bibinfo {volume} {20}},\ \bibinfo {pages}
  {103049} (\bibinfo {year} {2018})}\BibitemShut {NoStop}%
\bibitem [{\citenamefont {Bommer}\ \emph {et~al.}(2019)\citenamefont {Bommer},
  \citenamefont {Zhang}, \citenamefont {G{\"u}l}, \citenamefont {Nijholt},
  \citenamefont {Wimmer}, \citenamefont {Rybakov}, \citenamefont {Garaud},
  \citenamefont {Rodic}, \citenamefont {Babaev}, \citenamefont {Troyer} \emph
  {et~al.}}]{Jouri2019}%
  \BibitemOpen
  \bibfield  {author} {\bibinfo {author} {\bibfnamefont {J.~D.}\ \bibnamefont
  {Bommer}}, \bibinfo {author} {\bibfnamefont {H.}~\bibnamefont {Zhang}},
  \bibinfo {author} {\bibfnamefont {{\"O}.}~\bibnamefont {G{\"u}l}}, \bibinfo
  {author} {\bibfnamefont {B.}~\bibnamefont {Nijholt}}, \bibinfo {author}
  {\bibfnamefont {M.}~\bibnamefont {Wimmer}}, \bibinfo {author} {\bibfnamefont
  {F.~N.}\ \bibnamefont {Rybakov}}, \bibinfo {author} {\bibfnamefont
  {J.}~\bibnamefont {Garaud}}, \bibinfo {author} {\bibfnamefont
  {D.}~\bibnamefont {Rodic}}, \bibinfo {author} {\bibfnamefont
  {E.}~\bibnamefont {Babaev}}, \bibinfo {author} {\bibfnamefont
  {M.}~\bibnamefont {Troyer}}, \emph {et~al.},\ }\bibfield  {title} {\bibinfo
  {title} {Spin-orbit protection of induced superconductivity in majorana
  nanowires},\ }\href@noop {} {\bibfield  {journal} {\bibinfo  {journal}
  {Physical Review Letters}\ }\textbf {\bibinfo {volume} {122}},\ \bibinfo
  {pages} {187702} (\bibinfo {year} {2019})}\BibitemShut {NoStop}%
\bibitem [{\citenamefont {J{\"u}nger}\ \emph {et~al.}(2020)\citenamefont
  {J{\"u}nger}, \citenamefont {Delagrange}, \citenamefont {Chevallier},
  \citenamefont {Lehmann}, \citenamefont {Dick}, \citenamefont {Thelander},
  \citenamefont {Klinovaja}, \citenamefont {Loss}, \citenamefont
  {Baumgartner},\ and\ \citenamefont {Sch{\"o}nenberger}}]{Schonenberger2020}%
  \BibitemOpen
  \bibfield  {author} {\bibinfo {author} {\bibfnamefont {C.}~\bibnamefont
  {J{\"u}nger}}, \bibinfo {author} {\bibfnamefont {R.}~\bibnamefont
  {Delagrange}}, \bibinfo {author} {\bibfnamefont {D.}~\bibnamefont
  {Chevallier}}, \bibinfo {author} {\bibfnamefont {S.}~\bibnamefont {Lehmann}},
  \bibinfo {author} {\bibfnamefont {K.~A.}\ \bibnamefont {Dick}}, \bibinfo
  {author} {\bibfnamefont {C.}~\bibnamefont {Thelander}}, \bibinfo {author}
  {\bibfnamefont {J.}~\bibnamefont {Klinovaja}}, \bibinfo {author}
  {\bibfnamefont {D.}~\bibnamefont {Loss}}, \bibinfo {author} {\bibfnamefont
  {A.}~\bibnamefont {Baumgartner}},\ and\ \bibinfo {author} {\bibfnamefont
  {C.}~\bibnamefont {Sch{\"o}nenberger}},\ }\bibfield  {title} {\bibinfo
  {title} {Magnetic-field-independent subgap states in hybrid rashba
  nanowires},\ }\href@noop {} {\bibfield  {journal} {\bibinfo  {journal}
  {Physical Review Letters}\ }\textbf {\bibinfo {volume} {125}},\ \bibinfo
  {pages} {017701} (\bibinfo {year} {2020})}\BibitemShut {NoStop}%
\bibitem [{\citenamefont {Lutchyn}\ \emph {et~al.}(2010)\citenamefont
  {Lutchyn}, \citenamefont {Sau},\ and\ \citenamefont
  {Das~Sarma}}]{Lutchyn2010}%
  \BibitemOpen
  \bibfield  {author} {\bibinfo {author} {\bibfnamefont {R.~M.}\ \bibnamefont
  {Lutchyn}}, \bibinfo {author} {\bibfnamefont {J.~D.}\ \bibnamefont {Sau}},\
  and\ \bibinfo {author} {\bibfnamefont {S.}~\bibnamefont {Das~Sarma}},\
  }\bibfield  {title} {\bibinfo {title} {Majorana fermions and a topological
  phase transition in semiconductor-superconductor heterostructures},\ }\href
  {https://doi.org/10.1103/PhysRevLett.105.077001} {\bibfield  {journal}
  {\bibinfo  {journal} {Phys. Rev. Lett.}\ }\textbf {\bibinfo {volume} {105}},\
  \bibinfo {pages} {077001} (\bibinfo {year} {2010})}\BibitemShut {NoStop}%
\bibitem [{\citenamefont {Oreg}\ \emph {et~al.}(2010)\citenamefont {Oreg},
  \citenamefont {Refael},\ and\ \citenamefont {von Oppen}}]{Oreg2010}%
  \BibitemOpen
  \bibfield  {author} {\bibinfo {author} {\bibfnamefont {Y.}~\bibnamefont
  {Oreg}}, \bibinfo {author} {\bibfnamefont {G.}~\bibnamefont {Refael}},\ and\
  \bibinfo {author} {\bibfnamefont {F.}~\bibnamefont {von Oppen}},\ }\bibfield
  {title} {\bibinfo {title} {Helical liquids and majorana bound states in
  quantum wires},\ }\href {https://doi.org/10.1103/PhysRevLett.105.177002}
  {\bibfield  {journal} {\bibinfo  {journal} {Phys. Rev. Lett.}\ }\textbf
  {\bibinfo {volume} {105}},\ \bibinfo {pages} {177002} (\bibinfo {year}
  {2010})}\BibitemShut {NoStop}%
\bibitem [{\citenamefont {Mourik}\ \emph {et~al.}(2012)\citenamefont {Mourik},
  \citenamefont {Zuo}, \citenamefont {Frolov}, \citenamefont {Plissard},
  \citenamefont {Bakkers},\ and\ \citenamefont {Kouwenhoven}}]{Mourik}%
  \BibitemOpen
  \bibfield  {author} {\bibinfo {author} {\bibfnamefont {V.}~\bibnamefont
  {Mourik}}, \bibinfo {author} {\bibfnamefont {K.}~\bibnamefont {Zuo}},
  \bibinfo {author} {\bibfnamefont {S.~M.}\ \bibnamefont {Frolov}}, \bibinfo
  {author} {\bibfnamefont {S.}~\bibnamefont {Plissard}}, \bibinfo {author}
  {\bibfnamefont {E.~P.}\ \bibnamefont {Bakkers}},\ and\ \bibinfo {author}
  {\bibfnamefont {L.~P.}\ \bibnamefont {Kouwenhoven}},\ }\bibfield  {title}
  {\bibinfo {title} {Signatures of majorana fermions in hybrid
  superconductor-semiconductor nanowire devices},\ }\href@noop {} {\bibfield
  {journal} {\bibinfo  {journal} {Science}\ }\textbf {\bibinfo {volume}
  {336}},\ \bibinfo {pages} {1003} (\bibinfo {year} {2012})}\BibitemShut
  {NoStop}%
\bibitem [{\citenamefont {Deng}\ \emph {et~al.}(2016)\citenamefont {Deng},
  \citenamefont {Vaitiek{\.e}nas}, \citenamefont {Hansen}, \citenamefont
  {Danon}, \citenamefont {Leijnse}, \citenamefont {Flensberg}, \citenamefont
  {Nyg{\aa}rd}, \citenamefont {Krogstrup},\ and\ \citenamefont
  {Marcus}}]{Deng2016}%
  \BibitemOpen
  \bibfield  {author} {\bibinfo {author} {\bibfnamefont {M.}~\bibnamefont
  {Deng}}, \bibinfo {author} {\bibfnamefont {S.}~\bibnamefont
  {Vaitiek{\.e}nas}}, \bibinfo {author} {\bibfnamefont {E.~B.}\ \bibnamefont
  {Hansen}}, \bibinfo {author} {\bibfnamefont {J.}~\bibnamefont {Danon}},
  \bibinfo {author} {\bibfnamefont {M.}~\bibnamefont {Leijnse}}, \bibinfo
  {author} {\bibfnamefont {K.}~\bibnamefont {Flensberg}}, \bibinfo {author}
  {\bibfnamefont {J.}~\bibnamefont {Nyg{\aa}rd}}, \bibinfo {author}
  {\bibfnamefont {P.}~\bibnamefont {Krogstrup}},\ and\ \bibinfo {author}
  {\bibfnamefont {C.~M.}\ \bibnamefont {Marcus}},\ }\bibfield  {title}
  {\bibinfo {title} {Majorana bound state in a coupled quantum-dot
  hybrid-nanowire system},\ }\href@noop {} {\bibfield  {journal} {\bibinfo
  {journal} {Science}\ }\textbf {\bibinfo {volume} {354}},\ \bibinfo {pages}
  {1557} (\bibinfo {year} {2016})}\BibitemShut {NoStop}%
\bibitem [{\citenamefont {G{\"u}l}\ \emph {et~al.}(2018)\citenamefont
  {G{\"u}l}, \citenamefont {Zhang}, \citenamefont {Bommer}, \citenamefont
  {de~Moor}, \citenamefont {Car}, \citenamefont {Plissard}, \citenamefont
  {Bakkers}, \citenamefont {Geresdi}, \citenamefont {Watanabe}, \citenamefont
  {Taniguchi} \emph {et~al.}}]{Gul2018}%
  \BibitemOpen
  \bibfield  {author} {\bibinfo {author} {\bibfnamefont {{\"O}.}~\bibnamefont
  {G{\"u}l}}, \bibinfo {author} {\bibfnamefont {H.}~\bibnamefont {Zhang}},
  \bibinfo {author} {\bibfnamefont {J.~D.}\ \bibnamefont {Bommer}}, \bibinfo
  {author} {\bibfnamefont {M.~W.}\ \bibnamefont {de~Moor}}, \bibinfo {author}
  {\bibfnamefont {D.}~\bibnamefont {Car}}, \bibinfo {author} {\bibfnamefont
  {S.~R.}\ \bibnamefont {Plissard}}, \bibinfo {author} {\bibfnamefont {E.~P.}\
  \bibnamefont {Bakkers}}, \bibinfo {author} {\bibfnamefont {A.}~\bibnamefont
  {Geresdi}}, \bibinfo {author} {\bibfnamefont {K.}~\bibnamefont {Watanabe}},
  \bibinfo {author} {\bibfnamefont {T.}~\bibnamefont {Taniguchi}}, \emph
  {et~al.},\ }\bibfield  {title} {\bibinfo {title} {Ballistic majorana nanowire
  devices},\ }\href@noop {} {\bibfield  {journal} {\bibinfo  {journal} {Nature
  Nanotechnology}\ }\textbf {\bibinfo {volume} {13}},\ \bibinfo {pages} {192}
  (\bibinfo {year} {2018})}\BibitemShut {NoStop}%
\bibitem [{\citenamefont {Zhang}\ \emph {et~al.}(2021)\citenamefont {Zhang},
  \citenamefont {de~Moor}, \citenamefont {Bommer}, \citenamefont {Xu},
  \citenamefont {Wang}, \citenamefont {van Loo}, \citenamefont {Liu},
  \citenamefont {Gazibegovic}, \citenamefont {Logan}, \citenamefont {Car},
  \citenamefont {Op~het Veld}, \citenamefont {van Veldhoven}, \citenamefont
  {Koellinga}, \citenamefont {Verheijen}, \citenamefont {Pendharkar},
  \citenamefont {Pennachio}, \citenamefont {Shojaei}, \citenamefont {Lee},
  \citenamefont {Palmstr{\o}m}, \citenamefont {Bakkers}, \citenamefont
  {Das~Sarma},\ and\ \citenamefont {Kouwenhoven}}]{Zhang2021}%
  \BibitemOpen
  \bibfield  {author} {\bibinfo {author} {\bibfnamefont {H.}~\bibnamefont
  {Zhang}}, \bibinfo {author} {\bibfnamefont {M.~W.}\ \bibnamefont {de~Moor}},
  \bibinfo {author} {\bibfnamefont {J.~D.}\ \bibnamefont {Bommer}}, \bibinfo
  {author} {\bibfnamefont {D.}~\bibnamefont {Xu}}, \bibinfo {author}
  {\bibfnamefont {G.}~\bibnamefont {Wang}}, \bibinfo {author} {\bibfnamefont
  {N.}~\bibnamefont {van Loo}}, \bibinfo {author} {\bibfnamefont {C.-X.}\
  \bibnamefont {Liu}}, \bibinfo {author} {\bibfnamefont {S.}~\bibnamefont
  {Gazibegovic}}, \bibinfo {author} {\bibfnamefont {J.~A.}\ \bibnamefont
  {Logan}}, \bibinfo {author} {\bibfnamefont {D.}~\bibnamefont {Car}}, \bibinfo
  {author} {\bibfnamefont {R.~L.~M.}\ \bibnamefont {Op~het Veld}}, \bibinfo
  {author} {\bibfnamefont {P.~J.}\ \bibnamefont {van Veldhoven}}, \bibinfo
  {author} {\bibfnamefont {S.}~\bibnamefont {Koellinga}}, \bibinfo {author}
  {\bibfnamefont {M.~A.}\ \bibnamefont {Verheijen}}, \bibinfo {author}
  {\bibfnamefont {M.}~\bibnamefont {Pendharkar}}, \bibinfo {author}
  {\bibfnamefont {D.~J.}\ \bibnamefont {Pennachio}}, \bibinfo {author}
  {\bibfnamefont {B.}~\bibnamefont {Shojaei}}, \bibinfo {author} {\bibfnamefont
  {J.~S.}\ \bibnamefont {Lee}}, \bibinfo {author} {\bibfnamefont {C.~J.}\
  \bibnamefont {Palmstr{\o}m}}, \bibinfo {author} {\bibfnamefont {E.~P.}\
  \bibnamefont {Bakkers}}, \bibinfo {author} {\bibfnamefont {S.}~\bibnamefont
  {Das~Sarma}},\ and\ \bibinfo {author} {\bibfnamefont {L.~P.}\ \bibnamefont
  {Kouwenhoven}},\ }\bibfield  {title} {\bibinfo {title} {Large zero-bias peaks
  in insb-al hybrid semiconductor-superconductor nanowire devices},\
  }\href@noop {} {\bibfield  {journal} {\bibinfo  {journal} {arXiv:
  2101.11456}\ } (\bibinfo {year} {2021})}\BibitemShut {NoStop}%
\bibitem [{\citenamefont {Song}\ \emph {et~al.}(2021)\citenamefont {Song},
  \citenamefont {Zhang}, \citenamefont {Pan}, \citenamefont {Liu},
  \citenamefont {Wang}, \citenamefont {Cao}, \citenamefont {Liu}, \citenamefont
  {Wen}, \citenamefont {Liao}, \citenamefont {Zhuo}, \citenamefont {Liu},
  \citenamefont {Shang}, \citenamefont {Zhao},\ and\ \citenamefont
  {Zhang}}]{Song2021}%
  \BibitemOpen
  \bibfield  {author} {\bibinfo {author} {\bibfnamefont {H.}~\bibnamefont
  {Song}}, \bibinfo {author} {\bibfnamefont {Z.}~\bibnamefont {Zhang}},
  \bibinfo {author} {\bibfnamefont {D.}~\bibnamefont {Pan}}, \bibinfo {author}
  {\bibfnamefont {D.}~\bibnamefont {Liu}}, \bibinfo {author} {\bibfnamefont
  {Z.}~\bibnamefont {Wang}}, \bibinfo {author} {\bibfnamefont {Z.}~\bibnamefont
  {Cao}}, \bibinfo {author} {\bibfnamefont {L.}~\bibnamefont {Liu}}, \bibinfo
  {author} {\bibfnamefont {L.}~\bibnamefont {Wen}}, \bibinfo {author}
  {\bibfnamefont {D.}~\bibnamefont {Liao}}, \bibinfo {author} {\bibfnamefont
  {R.}~\bibnamefont {Zhuo}}, \bibinfo {author} {\bibfnamefont {D.~E.}\
  \bibnamefont {Liu}}, \bibinfo {author} {\bibfnamefont {R.}~\bibnamefont
  {Shang}}, \bibinfo {author} {\bibfnamefont {J.}~\bibnamefont {Zhao}},\ and\
  \bibinfo {author} {\bibfnamefont {H.}~\bibnamefont {Zhang}},\ }\bibfield
  {title} {\bibinfo {title} {Large zero bias peaks and dips in a four-terminal
  thin inas-al nanowire device},\ }\href@noop {} {\bibfield  {journal}
  {\bibinfo  {journal} {arXiv: 2107.08282}\ } (\bibinfo {year}
  {2021})}\BibitemShut {NoStop}%
\bibitem [{\citenamefont {Prada}\ \emph {et~al.}(2020)\citenamefont {Prada},
  \citenamefont {San-Jose}, \citenamefont {de~Moor}, \citenamefont {Geresdi},
  \citenamefont {Lee}, \citenamefont {Klinovaja}, \citenamefont {Loss},
  \citenamefont {Nyg{\aa}rd}, \citenamefont {Aguado},\ and\ \citenamefont
  {Kouwenhoven}}]{Prada2020}%
  \BibitemOpen
  \bibfield  {author} {\bibinfo {author} {\bibfnamefont {E.}~\bibnamefont
  {Prada}}, \bibinfo {author} {\bibfnamefont {P.}~\bibnamefont {San-Jose}},
  \bibinfo {author} {\bibfnamefont {M.~W.}\ \bibnamefont {de~Moor}}, \bibinfo
  {author} {\bibfnamefont {A.}~\bibnamefont {Geresdi}}, \bibinfo {author}
  {\bibfnamefont {E.~J.}\ \bibnamefont {Lee}}, \bibinfo {author} {\bibfnamefont
  {J.}~\bibnamefont {Klinovaja}}, \bibinfo {author} {\bibfnamefont
  {D.}~\bibnamefont {Loss}}, \bibinfo {author} {\bibfnamefont {J.}~\bibnamefont
  {Nyg{\aa}rd}}, \bibinfo {author} {\bibfnamefont {R.}~\bibnamefont {Aguado}},\
  and\ \bibinfo {author} {\bibfnamefont {L.~P.}\ \bibnamefont {Kouwenhoven}},\
  }\bibfield  {title} {\bibinfo {title} {From andreev to majorana bound states
  in hybrid superconductor--semiconductor nanowires},\ }\href@noop {}
  {\bibfield  {journal} {\bibinfo  {journal} {Nature Reviews Physics}\ }\textbf
  {\bibinfo {volume} {2}},\ \bibinfo {pages} {575} (\bibinfo {year}
  {2020})}\BibitemShut {NoStop}%
\bibitem [{\citenamefont {Prada}\ \emph {et~al.}(2012)\citenamefont {Prada},
  \citenamefont {San-Jose},\ and\ \citenamefont {Aguado}}]{Prada2012}%
  \BibitemOpen
  \bibfield  {author} {\bibinfo {author} {\bibfnamefont {E.}~\bibnamefont
  {Prada}}, \bibinfo {author} {\bibfnamefont {P.}~\bibnamefont {San-Jose}},\
  and\ \bibinfo {author} {\bibfnamefont {R.}~\bibnamefont {Aguado}},\
  }\bibfield  {title} {\bibinfo {title} {Transport spectroscopy of ns nanowire
  junctions with majorana fermions},\ }\href@noop {} {\bibfield  {journal}
  {\bibinfo  {journal} {Physical Review B}\ }\textbf {\bibinfo {volume} {86}},\
  \bibinfo {pages} {180503} (\bibinfo {year} {2012})}\BibitemShut {NoStop}%
\bibitem [{\citenamefont {Liu}\ \emph {et~al.}(2012)\citenamefont {Liu},
  \citenamefont {Potter}, \citenamefont {Law},\ and\ \citenamefont
  {Lee}}]{Patrick_Lee_disorder_2012}%
  \BibitemOpen
  \bibfield  {author} {\bibinfo {author} {\bibfnamefont {J.}~\bibnamefont
  {Liu}}, \bibinfo {author} {\bibfnamefont {A.~C.}\ \bibnamefont {Potter}},
  \bibinfo {author} {\bibfnamefont {K.~T.}\ \bibnamefont {Law}},\ and\ \bibinfo
  {author} {\bibfnamefont {P.~A.}\ \bibnamefont {Lee}},\ }\bibfield  {title}
  {\bibinfo {title} {Zero-bias peaks in the tunneling conductance of
  spin-orbit-coupled superconducting wires with and without majorana
  end-states},\ }\href {https://doi.org/10.1103/PhysRevLett.109.267002}
  {\bibfield  {journal} {\bibinfo  {journal} {Phys. Rev. Lett.}\ }\textbf
  {\bibinfo {volume} {109}},\ \bibinfo {pages} {267002} (\bibinfo {year}
  {2012})}\BibitemShut {NoStop}%
\bibitem [{\citenamefont {Kells}\ \emph {et~al.}(2012)\citenamefont {Kells},
  \citenamefont {Meidan},\ and\ \citenamefont {Brouwer}}]{BrouwerSmooth}%
  \BibitemOpen
  \bibfield  {author} {\bibinfo {author} {\bibfnamefont {G.}~\bibnamefont
  {Kells}}, \bibinfo {author} {\bibfnamefont {D.}~\bibnamefont {Meidan}},\ and\
  \bibinfo {author} {\bibfnamefont {P.}~\bibnamefont {Brouwer}},\ }\bibfield
  {title} {\bibinfo {title} {Near-zero-energy end states in topologically
  trivial spin-orbit coupled superconducting nanowires with a smooth
  confinement},\ }\href@noop {} {\bibfield  {journal} {\bibinfo  {journal}
  {Physical Review B}\ }\textbf {\bibinfo {volume} {86}},\ \bibinfo {pages}
  {100503} (\bibinfo {year} {2012})}\BibitemShut {NoStop}%
\bibitem [{\citenamefont {Liu}\ \emph {et~al.}(2017)\citenamefont {Liu},
  \citenamefont {Sau}, \citenamefont {Stanescu},\ and\ \citenamefont
  {Sarma}}]{Liu2017}%
  \BibitemOpen
  \bibfield  {author} {\bibinfo {author} {\bibfnamefont {C.-X.}\ \bibnamefont
  {Liu}}, \bibinfo {author} {\bibfnamefont {J.~D.}\ \bibnamefont {Sau}},
  \bibinfo {author} {\bibfnamefont {T.~D.}\ \bibnamefont {Stanescu}},\ and\
  \bibinfo {author} {\bibfnamefont {S.~D.}\ \bibnamefont {Sarma}},\ }\bibfield
  {title} {\bibinfo {title} {Andreev bound states versus majorana bound states
  in quantum dot-nanowire-superconductor hybrid structures: Trivial versus
  topological zero-bias conductance peaks},\ }\href@noop {} {\bibfield
  {journal} {\bibinfo  {journal} {Physical Review B}\ }\textbf {\bibinfo
  {volume} {96}},\ \bibinfo {pages} {075161} (\bibinfo {year}
  {2017})}\BibitemShut {NoStop}%
\bibitem [{\citenamefont {Moore}\ \emph {et~al.}(2018)\citenamefont {Moore},
  \citenamefont {Zeng}, \citenamefont {Stanescu},\ and\ \citenamefont
  {Tewari}}]{TudorQuasi}%
  \BibitemOpen
  \bibfield  {author} {\bibinfo {author} {\bibfnamefont {C.}~\bibnamefont
  {Moore}}, \bibinfo {author} {\bibfnamefont {C.}~\bibnamefont {Zeng}},
  \bibinfo {author} {\bibfnamefont {T.~D.}\ \bibnamefont {Stanescu}},\ and\
  \bibinfo {author} {\bibfnamefont {S.}~\bibnamefont {Tewari}},\ }\bibfield
  {title} {\bibinfo {title} {Quantized zero-bias conductance plateau in
  semiconductor-superconductor heterostructures without topological majorana
  zero modes},\ }\href@noop {} {\bibfield  {journal} {\bibinfo  {journal}
  {Physical Review B}\ }\textbf {\bibinfo {volume} {98}},\ \bibinfo {pages}
  {155314} (\bibinfo {year} {2018})}\BibitemShut {NoStop}%
\bibitem [{\citenamefont {Vuik}\ \emph {et~al.}(2019)\citenamefont {Vuik},
  \citenamefont {Nijholt}, \citenamefont {Akhmerov},\ and\ \citenamefont
  {Wimmer}}]{WimmerQuasi}%
  \BibitemOpen
  \bibfield  {author} {\bibinfo {author} {\bibfnamefont {A.}~\bibnamefont
  {Vuik}}, \bibinfo {author} {\bibfnamefont {B.}~\bibnamefont {Nijholt}},
  \bibinfo {author} {\bibfnamefont {A.}~\bibnamefont {Akhmerov}},\ and\
  \bibinfo {author} {\bibfnamefont {M.}~\bibnamefont {Wimmer}},\ }\bibfield
  {title} {\bibinfo {title} {Reproducing topological properties with
  quasi-majorana states},\ }\href@noop {} {\bibfield  {journal} {\bibinfo
  {journal} {SciPost Physics}\ }\textbf {\bibinfo {volume} {7}},\ \bibinfo
  {pages} {061} (\bibinfo {year} {2019})}\BibitemShut {NoStop}%
\bibitem [{\citenamefont {Liu}\ \emph {et~al.}(2018)\citenamefont {Liu},
  \citenamefont {Rossi},\ and\ \citenamefont {Lutchyn}}]{DEL-Disorder2018}%
  \BibitemOpen
  \bibfield  {author} {\bibinfo {author} {\bibfnamefont {D.~E.}\ \bibnamefont
  {Liu}}, \bibinfo {author} {\bibfnamefont {E.}~\bibnamefont {Rossi}},\ and\
  \bibinfo {author} {\bibfnamefont {R.~M.}\ \bibnamefont {Lutchyn}},\
  }\bibfield  {title} {\bibinfo {title} {Impurity-induced states in
  superconducting heterostructures},\ }\href
  {https://doi.org/10.1103/PhysRevB.97.161408} {\bibfield  {journal} {\bibinfo
  {journal} {Phys. Rev. B}\ }\textbf {\bibinfo {volume} {97}},\ \bibinfo
  {pages} {161408} (\bibinfo {year} {2018})}\BibitemShut {NoStop}%
\bibitem [{\citenamefont {Cao}\ \emph {et~al.}(2019)\citenamefont {Cao},
  \citenamefont {Zhang}, \citenamefont {L\"u}, \citenamefont {He},
  \citenamefont {Lu},\ and\ \citenamefont {Xie}}]{CaoZhanPRL}%
  \BibitemOpen
  \bibfield  {author} {\bibinfo {author} {\bibfnamefont {Z.}~\bibnamefont
  {Cao}}, \bibinfo {author} {\bibfnamefont {H.}~\bibnamefont {Zhang}}, \bibinfo
  {author} {\bibfnamefont {H.-F.}\ \bibnamefont {L\"u}}, \bibinfo {author}
  {\bibfnamefont {W.-X.}\ \bibnamefont {He}}, \bibinfo {author} {\bibfnamefont
  {H.-Z.}\ \bibnamefont {Lu}},\ and\ \bibinfo {author} {\bibfnamefont {X.~C.}\
  \bibnamefont {Xie}},\ }\bibfield  {title} {\bibinfo {title} {Decays of
  majorana or andreev oscillations induced by steplike spin-orbit coupling},\
  }\href {https://doi.org/10.1103/PhysRevLett.122.147701} {\bibfield  {journal}
  {\bibinfo  {journal} {Phys. Rev. Lett.}\ }\textbf {\bibinfo {volume} {122}},\
  \bibinfo {pages} {147701} (\bibinfo {year} {2019})}\BibitemShut {NoStop}%
\bibitem [{\citenamefont {Reeg}\ \emph {et~al.}(2018)\citenamefont {Reeg},
  \citenamefont {Dmytruk}, \citenamefont {Chevallier}, \citenamefont {Loss},\
  and\ \citenamefont {Klinovaja}}]{Loss2018ABS}%
  \BibitemOpen
  \bibfield  {author} {\bibinfo {author} {\bibfnamefont {C.}~\bibnamefont
  {Reeg}}, \bibinfo {author} {\bibfnamefont {O.}~\bibnamefont {Dmytruk}},
  \bibinfo {author} {\bibfnamefont {D.}~\bibnamefont {Chevallier}}, \bibinfo
  {author} {\bibfnamefont {D.}~\bibnamefont {Loss}},\ and\ \bibinfo {author}
  {\bibfnamefont {J.}~\bibnamefont {Klinovaja}},\ }\bibfield  {title} {\bibinfo
  {title} {Zero-energy andreev bound states from quantum dots in proximitized
  rashba nanowires},\ }\href@noop {} {\bibfield  {journal} {\bibinfo  {journal}
  {Physical Review B}\ }\textbf {\bibinfo {volume} {98}},\ \bibinfo {pages}
  {245407} (\bibinfo {year} {2018})}\BibitemShut {NoStop}%
\bibitem [{\citenamefont {Pan}\ and\ \citenamefont
  {Das~Sarma}(2020)}]{GoodBadUgly}%
  \BibitemOpen
  \bibfield  {author} {\bibinfo {author} {\bibfnamefont {H.}~\bibnamefont
  {Pan}}\ and\ \bibinfo {author} {\bibfnamefont {S.}~\bibnamefont
  {Das~Sarma}},\ }\bibfield  {title} {\bibinfo {title} {Physical mechanisms for
  zero-bias conductance peaks in majorana nanowires},\ }\href
  {https://doi.org/10.1103/PhysRevResearch.2.013377} {\bibfield  {journal}
  {\bibinfo  {journal} {Phys. Rev. Research}\ }\textbf {\bibinfo {volume}
  {2}},\ \bibinfo {pages} {013377} (\bibinfo {year} {2020})}\BibitemShut
  {NoStop}%
\bibitem [{\citenamefont {Das~Sarma}\ and\ \citenamefont
  {Pan}(2021)}]{DasSarma2021Disorder}%
  \BibitemOpen
  \bibfield  {author} {\bibinfo {author} {\bibfnamefont {S.}~\bibnamefont
  {Das~Sarma}}\ and\ \bibinfo {author} {\bibfnamefont {H.}~\bibnamefont
  {Pan}},\ }\bibfield  {title} {\bibinfo {title} {Disorder-induced zero-bias
  peaks in majorana nanowires},\ }\href
  {https://doi.org/10.1103/PhysRevB.103.195158} {\bibfield  {journal} {\bibinfo
   {journal} {Phys. Rev. B}\ }\textbf {\bibinfo {volume} {103}},\ \bibinfo
  {pages} {195158} (\bibinfo {year} {2021})}\BibitemShut {NoStop}%
\bibitem [{\citenamefont {Zeng}\ \emph {et~al.}(2021)\citenamefont {Zeng},
  \citenamefont {Sharma}, \citenamefont {Tewari},\ and\ \citenamefont
  {Stanescu}}]{Tudor2021Disorder}%
  \BibitemOpen
  \bibfield  {author} {\bibinfo {author} {\bibfnamefont {C.}~\bibnamefont
  {Zeng}}, \bibinfo {author} {\bibfnamefont {G.}~\bibnamefont {Sharma}},
  \bibinfo {author} {\bibfnamefont {S.}~\bibnamefont {Tewari}},\ and\ \bibinfo
  {author} {\bibfnamefont {T.}~\bibnamefont {Stanescu}},\ }\bibfield  {title}
  {\bibinfo {title} {Partially-separated majorana modes in a disordered
  medium},\ }\href@noop {} {\bibfield  {journal} {\bibinfo  {journal} {arXiv:
  2105.06469}\ } (\bibinfo {year} {2021})}\BibitemShut {NoStop}%
\bibitem [{\citenamefont {Liu}\ \emph {et~al.}(2021)\citenamefont {Liu},
  \citenamefont {Cao}, \citenamefont {Liu}, \citenamefont {Zhang},\ and\
  \citenamefont {Liu}}]{2021_PRB_Donghao}%
  \BibitemOpen
  \bibfield  {author} {\bibinfo {author} {\bibfnamefont {D.}~\bibnamefont
  {Liu}}, \bibinfo {author} {\bibfnamefont {Z.}~\bibnamefont {Cao}}, \bibinfo
  {author} {\bibfnamefont {X.}~\bibnamefont {Liu}}, \bibinfo {author}
  {\bibfnamefont {H.}~\bibnamefont {Zhang}},\ and\ \bibinfo {author}
  {\bibfnamefont {D.~E.}\ \bibnamefont {Liu}},\ }\bibfield  {title} {\bibinfo
  {title} {Topological kondo device for distinguishing quasi-majorana and
  majorana signatures},\ }\href {https://doi.org/10.1103/PhysRevB.104.205125}
  {\bibfield  {journal} {\bibinfo  {journal} {Phys. Rev. B}\ }\textbf {\bibinfo
  {volume} {104}},\ \bibinfo {pages} {205125} (\bibinfo {year}
  {2021})}\BibitemShut {NoStop}%
\bibitem [{\citenamefont {Liu}(2013)}]{Dong_PRL2013}%
  \BibitemOpen
  \bibfield  {author} {\bibinfo {author} {\bibfnamefont {D.~E.}\ \bibnamefont
  {Liu}},\ }\bibfield  {title} {\bibinfo {title} {Proposed method for tunneling
  spectroscopy with ohmic dissipation using resistive electrodes: a possible
  majorana filter},\ }\href {https://doi.org/10.1103/PhysRevLett.111.207003}
  {\bibfield  {journal} {\bibinfo  {journal} {Physical Review Letters}\
  }\textbf {\bibinfo {volume} {111}},\ \bibinfo {pages} {207003} (\bibinfo
  {year} {2013})}\BibitemShut {NoStop}%
\bibitem [{\citenamefont {Zhang}\ \emph {et~al.}(2022)\citenamefont {Zhang},
  \citenamefont {Wang}, \citenamefont {Pan}, \citenamefont {Li}, \citenamefont
  {Lu}, \citenamefont {Li}, \citenamefont {Zhang}, \citenamefont {Liu},
  \citenamefont {Cao}, \citenamefont {Liu}, \citenamefont {Wen}, \citenamefont
  {Liao}, \citenamefont {Zhuo}, \citenamefont {Shang}, \citenamefont {Liu},
  \citenamefont {Zhao},\ and\ \citenamefont {Zhang}}]{ZhangShan}%
  \BibitemOpen
  \bibfield  {author} {\bibinfo {author} {\bibfnamefont {S.}~\bibnamefont
  {Zhang}}, \bibinfo {author} {\bibfnamefont {Z.}~\bibnamefont {Wang}},
  \bibinfo {author} {\bibfnamefont {D.}~\bibnamefont {Pan}}, \bibinfo {author}
  {\bibfnamefont {H.}~\bibnamefont {Li}}, \bibinfo {author} {\bibfnamefont
  {S.}~\bibnamefont {Lu}}, \bibinfo {author} {\bibfnamefont {Z.}~\bibnamefont
  {Li}}, \bibinfo {author} {\bibfnamefont {G.}~\bibnamefont {Zhang}}, \bibinfo
  {author} {\bibfnamefont {D.}~\bibnamefont {Liu}}, \bibinfo {author}
  {\bibfnamefont {Z.}~\bibnamefont {Cao}}, \bibinfo {author} {\bibfnamefont
  {L.}~\bibnamefont {Liu}}, \bibinfo {author} {\bibfnamefont {L.}~\bibnamefont
  {Wen}}, \bibinfo {author} {\bibfnamefont {D.}~\bibnamefont {Liao}}, \bibinfo
  {author} {\bibfnamefont {R.}~\bibnamefont {Zhuo}}, \bibinfo {author}
  {\bibfnamefont {R.}~\bibnamefont {Shang}}, \bibinfo {author} {\bibfnamefont
  {D.~E.}\ \bibnamefont {Liu}}, \bibinfo {author} {\bibfnamefont
  {J.}~\bibnamefont {Zhao}},\ and\ \bibinfo {author} {\bibfnamefont
  {H.}~\bibnamefont {Zhang}},\ }\bibfield  {title} {\bibinfo {title}
  {Suppressing andreev bound state zero bias peaks using a strongly dissipative
  lead},\ }\href {https://doi.org/10.1103/PhysRevLett.128.076803} {\bibfield
  {journal} {\bibinfo  {journal} {Phys. Rev. Lett.}\ }\textbf {\bibinfo
  {volume} {128}},\ \bibinfo {pages} {076803} (\bibinfo {year}
  {2022})}\BibitemShut {NoStop}%
\bibitem [{\citenamefont {Delsing}\ \emph {et~al.}(1989)\citenamefont
  {Delsing}, \citenamefont {Likharev}, \citenamefont {Kuzmin},\ and\
  \citenamefont {Claeson}}]{Delsing_1989}%
  \BibitemOpen
  \bibfield  {author} {\bibinfo {author} {\bibfnamefont {P.}~\bibnamefont
  {Delsing}}, \bibinfo {author} {\bibfnamefont {K.~K.}\ \bibnamefont
  {Likharev}}, \bibinfo {author} {\bibfnamefont {L.~S.}\ \bibnamefont
  {Kuzmin}},\ and\ \bibinfo {author} {\bibfnamefont {T.}~\bibnamefont
  {Claeson}},\ }\bibfield  {title} {\bibinfo {title} {Effect of high-frequency
  electrodynamic environment on the single-electron tunneling in ultrasmall
  junctions},\ }\href {https://doi.org/10.1103/PhysRevLett.63.1180} {\bibfield
  {journal} {\bibinfo  {journal} {Phys. Rev. Lett.}\ }\textbf {\bibinfo
  {volume} {63}},\ \bibinfo {pages} {1180} (\bibinfo {year}
  {1989})}\BibitemShut {NoStop}%
\bibitem [{\citenamefont {Ingold}\ and\ \citenamefont {Nazarov}()}]{Ingold}%
  \BibitemOpen
  \bibfield  {author} {\bibinfo {author} {\bibfnamefont {G.~L.}\ \bibnamefont
  {Ingold}}\ and\ \bibinfo {author} {\bibfnamefont {Y.~V.}\ \bibnamefont
  {Nazarov}},\ }\href@noop {} {\emph {\bibinfo {title} {Single Charge
  Tunnelling: Coulomb Blockade Phenomena in Nanostructures}}},\ \bibinfo {note}
  {edited by H. Grabert and M. H. Devoret (Springer, New York, 1992), pp.
  21-107}\BibitemShut {NoStop}%
\bibitem [{\citenamefont {Flensberg}\ \emph {et~al.}(1992)\citenamefont
  {Flensberg}, \citenamefont {Girvin}, \citenamefont {Jonson}, \citenamefont
  {Penn},\ and\ \citenamefont {Stiles}}]{Flensberg_1992}%
  \BibitemOpen
  \bibfield  {author} {\bibinfo {author} {\bibfnamefont {K.}~\bibnamefont
  {Flensberg}}, \bibinfo {author} {\bibfnamefont {S.~M.}\ \bibnamefont
  {Girvin}}, \bibinfo {author} {\bibfnamefont {M.}~\bibnamefont {Jonson}},
  \bibinfo {author} {\bibfnamefont {D.~R.}\ \bibnamefont {Penn}},\ and\
  \bibinfo {author} {\bibfnamefont {M.~D.}\ \bibnamefont {Stiles}},\ }\bibfield
   {title} {\bibinfo {title} {Quantum mechanics of the electromagnetic
  environment in the single-junction coulomb blockade},\ }\href
  {https://doi.org/10.1088/0031-8949/1992/t42/032} {\bibfield  {journal}
  {\bibinfo  {journal} {Physica Scripta}\ }\textbf {\bibinfo {volume} {T42}},\
  \bibinfo {pages} {189} (\bibinfo {year} {1992})}\BibitemShut {NoStop}%
\bibitem [{\citenamefont {Joyez}\ \emph {et~al.}(1998)\citenamefont {Joyez},
  \citenamefont {Esteve},\ and\ \citenamefont {Devoret}}]{Joyez_1998}%
  \BibitemOpen
  \bibfield  {author} {\bibinfo {author} {\bibfnamefont {P.}~\bibnamefont
  {Joyez}}, \bibinfo {author} {\bibfnamefont {D.}~\bibnamefont {Esteve}},\ and\
  \bibinfo {author} {\bibfnamefont {M.~H.}\ \bibnamefont {Devoret}},\
  }\bibfield  {title} {\bibinfo {title} {How is the coulomb blockade suppressed
  in high-conductance tunnel junctions?},\ }\href
  {https://doi.org/10.1103/PhysRevLett.80.1956} {\bibfield  {journal} {\bibinfo
   {journal} {Phys. Rev. Lett.}\ }\textbf {\bibinfo {volume} {80}},\ \bibinfo
  {pages} {1956} (\bibinfo {year} {1998})}\BibitemShut {NoStop}%
\bibitem [{\citenamefont {Zheng}\ \emph {et~al.}(1998)\citenamefont {Zheng},
  \citenamefont {Friedman}, \citenamefont {Averin}, \citenamefont {Han},\ and\
  \citenamefont {Lukens}}]{Zheng_1998}%
  \BibitemOpen
  \bibfield  {author} {\bibinfo {author} {\bibfnamefont {W.}~\bibnamefont
  {Zheng}}, \bibinfo {author} {\bibfnamefont {J.}~\bibnamefont {Friedman}},
  \bibinfo {author} {\bibfnamefont {D.}~\bibnamefont {Averin}}, \bibinfo
  {author} {\bibfnamefont {S.}~\bibnamefont {Han}},\ and\ \bibinfo {author}
  {\bibfnamefont {J.}~\bibnamefont {Lukens}},\ }\bibfield  {title} {\bibinfo
  {title} {Observation of strong coulomb blockade in resistively isolated
  tunnel junctions},\ }\href
  {https://doi.org/https://doi.org/10.1016/S0038-1098(98)00439-6} {\bibfield
  {journal} {\bibinfo  {journal} {Solid State Communications}\ }\textbf
  {\bibinfo {volume} {108}},\ \bibinfo {pages} {839} (\bibinfo {year}
  {1998})}\BibitemShut {NoStop}%
\bibitem [{\citenamefont {Parmentier}\ \emph {et~al.}(2011)\citenamefont
  {Parmentier}, \citenamefont {Anthore}, \citenamefont {Jezouin}, \citenamefont
  {Le~Sueur}, \citenamefont {Gennser}, \citenamefont {Cavanna}, \citenamefont
  {Mailly},\ and\ \citenamefont {Pierre}}]{Pierre2011}%
  \BibitemOpen
  \bibfield  {author} {\bibinfo {author} {\bibfnamefont {F.}~\bibnamefont
  {Parmentier}}, \bibinfo {author} {\bibfnamefont {A.}~\bibnamefont {Anthore}},
  \bibinfo {author} {\bibfnamefont {S.}~\bibnamefont {Jezouin}}, \bibinfo
  {author} {\bibfnamefont {H.}~\bibnamefont {Le~Sueur}}, \bibinfo {author}
  {\bibfnamefont {U.}~\bibnamefont {Gennser}}, \bibinfo {author} {\bibfnamefont
  {A.}~\bibnamefont {Cavanna}}, \bibinfo {author} {\bibfnamefont
  {D.}~\bibnamefont {Mailly}},\ and\ \bibinfo {author} {\bibfnamefont
  {F.}~\bibnamefont {Pierre}},\ }\bibfield  {title} {\bibinfo {title} {Strong
  back-action of a linear circuit on a single electronic quantum channel},\
  }\href@noop {} {\bibfield  {journal} {\bibinfo  {journal} {Nature Physics}\
  }\textbf {\bibinfo {volume} {7}},\ \bibinfo {pages} {935} (\bibinfo {year}
  {2011})}\BibitemShut {NoStop}%
\bibitem [{\citenamefont {Mebrahtu}\ \emph {et~al.}(2012)\citenamefont
  {Mebrahtu}, \citenamefont {Borzenets}, \citenamefont {Liu}, \citenamefont
  {Zheng}, \citenamefont {Bomze}, \citenamefont {Smirnov}, \citenamefont
  {Baranger},\ and\ \citenamefont {Finkelstein}}]{Gleb_Nature}%
  \BibitemOpen
  \bibfield  {author} {\bibinfo {author} {\bibfnamefont {H.~T.}\ \bibnamefont
  {Mebrahtu}}, \bibinfo {author} {\bibfnamefont {I.~V.}\ \bibnamefont
  {Borzenets}}, \bibinfo {author} {\bibfnamefont {D.~E.}\ \bibnamefont {Liu}},
  \bibinfo {author} {\bibfnamefont {H.}~\bibnamefont {Zheng}}, \bibinfo
  {author} {\bibfnamefont {Y.~V.}\ \bibnamefont {Bomze}}, \bibinfo {author}
  {\bibfnamefont {A.~I.}\ \bibnamefont {Smirnov}}, \bibinfo {author}
  {\bibfnamefont {H.~U.}\ \bibnamefont {Baranger}},\ and\ \bibinfo {author}
  {\bibfnamefont {G.}~\bibnamefont {Finkelstein}},\ }\bibfield  {title}
  {\bibinfo {title} {Quantum phase transition in a resonant level coupled to
  interacting leads},\ }\href@noop {} {\bibfield  {journal} {\bibinfo
  {journal} {Nature}\ }\textbf {\bibinfo {volume} {488}},\ \bibinfo {pages}
  {61} (\bibinfo {year} {2012})}\BibitemShut {NoStop}%
\bibitem [{\citenamefont {Mebrahtu}\ \emph {et~al.}(2013)\citenamefont
  {Mebrahtu}, \citenamefont {Borzenets}, \citenamefont {Zheng}, \citenamefont
  {Bomze}, \citenamefont {Smirnov}, \citenamefont {Florens}, \citenamefont
  {Baranger},\ and\ \citenamefont {Finkelstein}}]{Gleb_NaturePhysics}%
  \BibitemOpen
  \bibfield  {author} {\bibinfo {author} {\bibfnamefont {H.}~\bibnamefont
  {Mebrahtu}}, \bibinfo {author} {\bibfnamefont {I.}~\bibnamefont {Borzenets}},
  \bibinfo {author} {\bibfnamefont {H.}~\bibnamefont {Zheng}}, \bibinfo
  {author} {\bibfnamefont {Y.~V.}\ \bibnamefont {Bomze}}, \bibinfo {author}
  {\bibfnamefont {A.}~\bibnamefont {Smirnov}}, \bibinfo {author} {\bibfnamefont
  {S.}~\bibnamefont {Florens}}, \bibinfo {author} {\bibfnamefont
  {H.}~\bibnamefont {Baranger}},\ and\ \bibinfo {author} {\bibfnamefont
  {G.}~\bibnamefont {Finkelstein}},\ }\bibfield  {title} {\bibinfo {title}
  {Observation of majorana quantum critical behaviour in a resonant level
  coupled to a dissipative environment},\ }\href@noop {} {\bibfield  {journal}
  {\bibinfo  {journal} {Nature Physics}\ }\textbf {\bibinfo {volume} {9}},\
  \bibinfo {pages} {732} (\bibinfo {year} {2013})}\BibitemShut {NoStop}%
\bibitem [{\citenamefont {Liu}\ \emph {et~al.}(2014)\citenamefont {Liu},
  \citenamefont {Zheng}, \citenamefont {Finkelstein},\ and\ \citenamefont
  {Baranger}}]{Dong_PRB2014}%
  \BibitemOpen
  \bibfield  {author} {\bibinfo {author} {\bibfnamefont {D.~E.}\ \bibnamefont
  {Liu}}, \bibinfo {author} {\bibfnamefont {H.}~\bibnamefont {Zheng}}, \bibinfo
  {author} {\bibfnamefont {G.}~\bibnamefont {Finkelstein}},\ and\ \bibinfo
  {author} {\bibfnamefont {H.~U.}\ \bibnamefont {Baranger}},\ }\bibfield
  {title} {\bibinfo {title} {Tunable quantum phase transitions in a resonant
  level coupled to two dissipative baths},\ }\href
  {https://doi.org/10.1103/PhysRevB.89.085116} {\bibfield  {journal} {\bibinfo
  {journal} {Physical Review B}\ }\textbf {\bibinfo {volume} {89}},\ \bibinfo
  {pages} {085116} (\bibinfo {year} {2014})}\BibitemShut {NoStop}%
\bibitem [{\citenamefont {Jezouin}\ \emph {et~al.}(2013)\citenamefont
  {Jezouin}, \citenamefont {Albert}, \citenamefont {Parmentier}, \citenamefont
  {Anthore}, \citenamefont {Gennser}, \citenamefont {Cavanna}, \citenamefont
  {Safi},\ and\ \citenamefont {Pierre}}]{Jezouin_2013}%
  \BibitemOpen
  \bibfield  {author} {\bibinfo {author} {\bibfnamefont {S.}~\bibnamefont
  {Jezouin}}, \bibinfo {author} {\bibfnamefont {M.}~\bibnamefont {Albert}},
  \bibinfo {author} {\bibfnamefont {F.}~\bibnamefont {Parmentier}}, \bibinfo
  {author} {\bibfnamefont {A.}~\bibnamefont {Anthore}}, \bibinfo {author}
  {\bibfnamefont {U.}~\bibnamefont {Gennser}}, \bibinfo {author} {\bibfnamefont
  {A.}~\bibnamefont {Cavanna}}, \bibinfo {author} {\bibfnamefont
  {I.}~\bibnamefont {Safi}},\ and\ \bibinfo {author} {\bibfnamefont
  {F.}~\bibnamefont {Pierre}},\ }\bibfield  {title} {\bibinfo {title}
  {Tomonaga--luttinger physics in electronic quantum circuits},\ }\href@noop {}
  {\bibfield  {journal} {\bibinfo  {journal} {Nature Communications}\ }\textbf
  {\bibinfo {volume} {4}},\ \bibinfo {pages} {1802} (\bibinfo {year}
  {2013})}\BibitemShut {NoStop}%
\bibitem [{\citenamefont {Anthore}\ \emph {et~al.}(2018)\citenamefont
  {Anthore}, \citenamefont {Iftikhar}, \citenamefont {Boulat}, \citenamefont
  {Parmentier}, \citenamefont {Cavanna}, \citenamefont {Ouerghi}, \citenamefont
  {Gennser},\ and\ \citenamefont {Pierre}}]{Pierre_PRX}%
  \BibitemOpen
  \bibfield  {author} {\bibinfo {author} {\bibfnamefont {A.}~\bibnamefont
  {Anthore}}, \bibinfo {author} {\bibfnamefont {Z.}~\bibnamefont {Iftikhar}},
  \bibinfo {author} {\bibfnamefont {E.}~\bibnamefont {Boulat}}, \bibinfo
  {author} {\bibfnamefont {F.~D.}\ \bibnamefont {Parmentier}}, \bibinfo
  {author} {\bibfnamefont {A.}~\bibnamefont {Cavanna}}, \bibinfo {author}
  {\bibfnamefont {A.}~\bibnamefont {Ouerghi}}, \bibinfo {author} {\bibfnamefont
  {U.}~\bibnamefont {Gennser}},\ and\ \bibinfo {author} {\bibfnamefont
  {F.}~\bibnamefont {Pierre}},\ }\bibfield  {title} {\bibinfo {title} {Circuit
  quantum simulation of a tomonaga-luttinger liquid with an impurity},\ }\href
  {https://doi.org/10.1103/PhysRevX.8.031075} {\bibfield  {journal} {\bibinfo
  {journal} {Phys. Rev. X}\ }\textbf {\bibinfo {volume} {8}},\ \bibinfo {pages}
  {031075} (\bibinfo {year} {2018})}\BibitemShut {NoStop}%
\bibitem [{\citenamefont {Safi}\ and\ \citenamefont
  {Saleur}(2004)}]{Safi_2004}%
  \BibitemOpen
  \bibfield  {author} {\bibinfo {author} {\bibfnamefont {I.}~\bibnamefont
  {Safi}}\ and\ \bibinfo {author} {\bibfnamefont {H.}~\bibnamefont {Saleur}},\
  }\bibfield  {title} {\bibinfo {title} {One-channel conductor in an ohmic
  environment: Mapping to a tomonaga-luttinger liquid and full counting
  statistics},\ }\href {https://doi.org/10.1103/PhysRevLett.93.126602}
  {\bibfield  {journal} {\bibinfo  {journal} {Phys. Rev. Lett.}\ }\textbf
  {\bibinfo {volume} {93}},\ \bibinfo {pages} {126602} (\bibinfo {year}
  {2004})}\BibitemShut {NoStop}%
\bibitem [{\citenamefont {Pan}\ \emph {et~al.}(2020)\citenamefont {Pan},
  \citenamefont {Song}, \citenamefont {Zhang}, \citenamefont {Liu},
  \citenamefont {Wen}, \citenamefont {Liao}, \citenamefont {Zhuo},
  \citenamefont {Wang}, \citenamefont {Zhang}, \citenamefont {Yang},
  \citenamefont {Ying}, \citenamefont {Miao}, \citenamefont {Shang},
  \citenamefont {Zhang},\ and\ \citenamefont {Zhao}}]{Pan2020}%
  \BibitemOpen
  \bibfield  {author} {\bibinfo {author} {\bibfnamefont {D.}~\bibnamefont
  {Pan}}, \bibinfo {author} {\bibfnamefont {H.}~\bibnamefont {Song}}, \bibinfo
  {author} {\bibfnamefont {S.}~\bibnamefont {Zhang}}, \bibinfo {author}
  {\bibfnamefont {L.}~\bibnamefont {Liu}}, \bibinfo {author} {\bibfnamefont
  {L.}~\bibnamefont {Wen}}, \bibinfo {author} {\bibfnamefont {D.}~\bibnamefont
  {Liao}}, \bibinfo {author} {\bibfnamefont {R.}~\bibnamefont {Zhuo}}, \bibinfo
  {author} {\bibfnamefont {Z.}~\bibnamefont {Wang}}, \bibinfo {author}
  {\bibfnamefont {Z.}~\bibnamefont {Zhang}}, \bibinfo {author} {\bibfnamefont
  {S.}~\bibnamefont {Yang}}, \bibinfo {author} {\bibfnamefont {J.}~\bibnamefont
  {Ying}}, \bibinfo {author} {\bibfnamefont {W.}~\bibnamefont {Miao}}, \bibinfo
  {author} {\bibfnamefont {R.}~\bibnamefont {Shang}}, \bibinfo {author}
  {\bibfnamefont {H.}~\bibnamefont {Zhang}},\ and\ \bibinfo {author}
  {\bibfnamefont {J.}~\bibnamefont {Zhao}},\ }\bibfield  {title} {\bibinfo
  {title} {In situ epitaxy of pure phase ultra-thin inas-al nanowires for
  quantum devices},\ }\href@noop {} {\bibfield  {journal} {\bibinfo  {journal}
  {arXiv: 2011.13620}\ } (\bibinfo {year} {2020})}\BibitemShut {NoStop}%
\bibitem [{\citenamefont {Liu}\ \emph {et~al.}(2022)\citenamefont {Liu},
  \citenamefont {Zhang}, \citenamefont {Cao}, \citenamefont {Zhang},\ and\
  \citenamefont {Liu}}]{Dong_2021}%
  \BibitemOpen
  \bibfield  {author} {\bibinfo {author} {\bibfnamefont {D.}~\bibnamefont
  {Liu}}, \bibinfo {author} {\bibfnamefont {G.}~\bibnamefont {Zhang}}, \bibinfo
  {author} {\bibfnamefont {Z.}~\bibnamefont {Cao}}, \bibinfo {author}
  {\bibfnamefont {H.}~\bibnamefont {Zhang}},\ and\ \bibinfo {author}
  {\bibfnamefont {D.~E.}\ \bibnamefont {Liu}},\ }\bibfield  {title} {\bibinfo
  {title} {Universal conductance scaling of andreev reflections using a
  dissipative probe},\ }\href {https://doi.org/10.1103/PhysRevLett.128.076802}
  {\bibfield  {journal} {\bibinfo  {journal} {Phys. Rev. Lett.}\ }\textbf
  {\bibinfo {volume} {128}},\ \bibinfo {pages} {076802} (\bibinfo {year}
  {2022})}\BibitemShut {NoStop}%
\bibitem [{\citenamefont {Zhang}\ \emph {et~al.}(2019)\citenamefont {Zhang},
  \citenamefont {Liu}, \citenamefont {Wimmer},\ and\ \citenamefont
  {Kouwenhoven}}]{NextSteps}%
  \BibitemOpen
  \bibfield  {author} {\bibinfo {author} {\bibfnamefont {H.}~\bibnamefont
  {Zhang}}, \bibinfo {author} {\bibfnamefont {D.~E.}\ \bibnamefont {Liu}},
  \bibinfo {author} {\bibfnamefont {M.}~\bibnamefont {Wimmer}},\ and\ \bibinfo
  {author} {\bibfnamefont {L.~P.}\ \bibnamefont {Kouwenhoven}},\ }\bibfield
  {title} {\bibinfo {title} {Next steps of quantum transport in majorana
  nanowire devices},\ }\href@noop {} {\bibfield  {journal} {\bibinfo  {journal}
  {Nature Communications}\ }\textbf {\bibinfo {volume} {10}},\ \bibinfo {pages}
  {5128} (\bibinfo {year} {2019})}\BibitemShut {NoStop}%
\bibitem [{\citenamefont {Zhang}\ and\ \citenamefont
  {Sp{\aa}nsl{\"a}tt}(2020)}]{Gu_2020}%
  \BibitemOpen
  \bibfield  {author} {\bibinfo {author} {\bibfnamefont {G.}~\bibnamefont
  {Zhang}}\ and\ \bibinfo {author} {\bibfnamefont {C.}~\bibnamefont
  {Sp{\aa}nsl{\"a}tt}},\ }\bibfield  {title} {\bibinfo {title} {Distinguishing
  between topological and quasi majorana zero modes with a dissipative resonant
  level},\ }\href {https://doi.org/10.1103/PhysRevB.102.045111} {\bibfield
  {journal} {\bibinfo  {journal} {Physical Review B}\ }\textbf {\bibinfo
  {volume} {102}},\ \bibinfo {pages} {045111} (\bibinfo {year}
  {2020})}\BibitemShut {NoStop}%
\bibitem [{\citenamefont {Liu}\ \emph {et~al.}(2020)\citenamefont {Liu},
  \citenamefont {Cao}, \citenamefont {Zhang},\ and\ \citenamefont
  {Liu}}]{Dong_PRB2020}%
  \BibitemOpen
  \bibfield  {author} {\bibinfo {author} {\bibfnamefont {D.}~\bibnamefont
  {Liu}}, \bibinfo {author} {\bibfnamefont {Z.}~\bibnamefont {Cao}}, \bibinfo
  {author} {\bibfnamefont {H.}~\bibnamefont {Zhang}},\ and\ \bibinfo {author}
  {\bibfnamefont {D.~E.}\ \bibnamefont {Liu}},\ }\bibfield  {title} {\bibinfo
  {title} {Revealing the nonlocal coherent nature of majorana devices from
  dissipative teleportation},\ }\href
  {https://doi.org/10.1103/PhysRevB.101.081406} {\bibfield  {journal} {\bibinfo
   {journal} {Physical Review B}\ }\textbf {\bibinfo {volume} {101}},\ \bibinfo
  {pages} {081406} (\bibinfo {year} {2020})}\BibitemShut {NoStop}%
\bibitem [{\citenamefont {Sengupta}\ \emph {et~al.}(2001)\citenamefont
  {Sengupta}, \citenamefont {{\v{Z}}uti{\'c}}, \citenamefont {Kwon},
  \citenamefont {Yakovenko},\ and\ \citenamefont {Sarma}}]{DasSarma2001}%
  \BibitemOpen
  \bibfield  {author} {\bibinfo {author} {\bibfnamefont {K.}~\bibnamefont
  {Sengupta}}, \bibinfo {author} {\bibfnamefont {I.}~\bibnamefont
  {{\v{Z}}uti{\'c}}}, \bibinfo {author} {\bibfnamefont {H.-J.}\ \bibnamefont
  {Kwon}}, \bibinfo {author} {\bibfnamefont {V.~M.}\ \bibnamefont
  {Yakovenko}},\ and\ \bibinfo {author} {\bibfnamefont {S.~D.}\ \bibnamefont
  {Sarma}},\ }\bibfield  {title} {\bibinfo {title} {Midgap edge states and
  pairing symmetry of quasi-one-dimensional organic superconductors},\
  }\href@noop {} {\bibfield  {journal} {\bibinfo  {journal} {Physical Review
  B}\ }\textbf {\bibinfo {volume} {63}},\ \bibinfo {pages} {144531} (\bibinfo
  {year} {2001})}\BibitemShut {NoStop}%
\bibitem [{\citenamefont {Law}\ \emph {et~al.}(2009)\citenamefont {Law},
  \citenamefont {Lee},\ and\ \citenamefont {Ng}}]{Law2009}%
  \BibitemOpen
  \bibfield  {author} {\bibinfo {author} {\bibfnamefont {K.~T.}\ \bibnamefont
  {Law}}, \bibinfo {author} {\bibfnamefont {P.~A.}\ \bibnamefont {Lee}},\ and\
  \bibinfo {author} {\bibfnamefont {T.~K.}\ \bibnamefont {Ng}},\ }\bibfield
  {title} {\bibinfo {title} {Majorana fermion induced resonant andreev
  reflection},\ }\href@noop {} {\bibfield  {journal} {\bibinfo  {journal}
  {Physical Review Letters}\ }\textbf {\bibinfo {volume} {103}},\ \bibinfo
  {pages} {237001} (\bibinfo {year} {2009})}\BibitemShut {NoStop}%
\end{thebibliography}%

\newpage
\onecolumngrid


\section*{Supplementary material}

# Supplementary Information for "Large Andreev Bound State Zero Bias Peaks in a Weakly Dissipative Environment"

Zhichuan Wang,[1,2,3,*] Shan Zhang,[2,*] Dong Pan,[4,*] Gu Zhang,[5,*] Zezhou Xia,[2] Zonglin
Li,[2] Donghao Liu,[2] Zhan Cao,[5] Lei Liu,[4] Lianjun Wen,[4] Dunyuan Liao,[4] Ran Zhuo,[4]
Yongqing Li,[1,3] Dong E. Liu,[2,5,6] Runan Shang,[5,†] Jianhua Zhao,[4,‡] and Hao Zhang[2,5,6,§]

[1]*Beijing National Laboratory for Condensed Matter Physics,*
*Institute of Physics, Chinese Academy of Sciences, Beijing 100190, China*
[2]*State Key Laboratory of Low Dimensional Quantum Physics,*
*Department of Physics, Tsinghua University, Beijing 100084, China*
[3]*School of Physical Sciences, University of Chinese Academy of Sciences, Beijing 100049, China*
[4]*State Key Laboratory of Superlattices and Microstructures, Institute of Semiconductors,*
*Chinese Academy of Sciences, P. O. Box 912, Beijing 100083, China*
[5]*Beijing Academy of Quantum Information Sciences, 100193 Beijing, China*
[6]*Frontier Science Center for Quantum Information, 100084 Beijing, China*

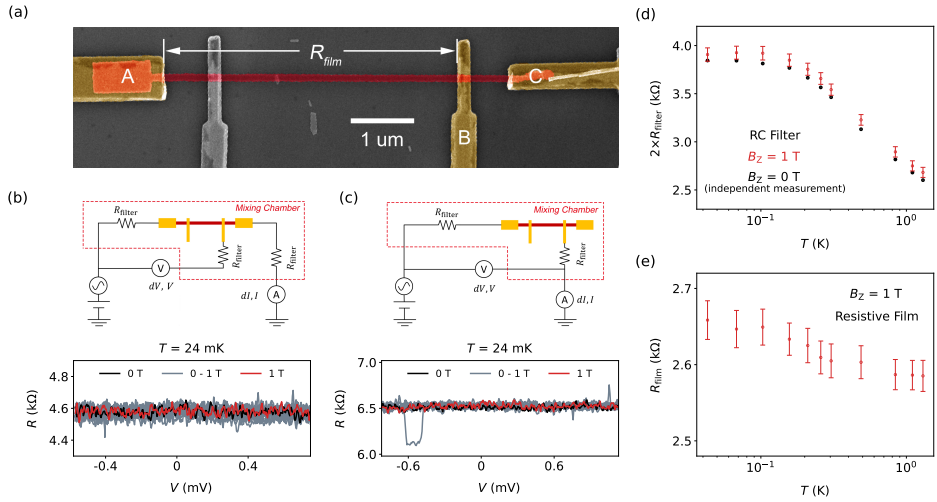

Fig. S 1: Dissipative film and fridge filter calibration. (a) False-color SEM of a Cr/Au dissipative film (red)
fabricated together with the dissipative resistor of Device A and B on the same chip. The gray Ti/Au contact was
broken and left floated, and the other three (labeled A, B, C) were used for calibration. (b) Upper, the
three-terminal circuit: a current $I$ flows from A to C while the voltage $V$ is measured between A and B. Each fridge
line has an RC-filter on the mixing chamber (see $R_{\text{filter}}$, capacitors not shown). Lower, differential resistance
$R = dV/dI$ shows no obvious dependency on $V$ and $B$. (c) A two-terminal circuit with C floated. A small instability
jump happened for one gray curve. $R$ in (b) is equal to $R_{\text{film}} + R_{\text{filter}}$. $R$ in (c) is equal to $R_{\text{film}} + 2R_{\text{filters}} + R_{\text{contact}}$,
where $R_{\text{contact}}$ is the contact resistance between B and the film. We plot the difference between (b) and (c), i.e.
$R_{\text{filter}} + R_{\text{contact}}$, as red dots ($\times 2$) in (d) for different $T$s. The error bars are $\sim 50$ - 70 $\Omega$, estimated based on the
fluctuations in (b-c). The black dots in (d) are independent calibration of two filters, measured by shorting two
fridge lines at the mixing chamber. The back dots and red dots show reasonable match with a small difference of $\sim$
70 $\Omega$, probably contributed by $2R_{\text{contact}}$. (e) $T$ dependence of $R_{\text{film}}$, calculated by subtracting $R_{\text{filter}}$ (black dots in
(d) divided by 2) from the $R$ in (b). Square resistance is then extracted and further used to estimate the resistance
of the dissipative resistor for Device A and B. $G$ in this work refers to the InAs-Al part where the resistances of the
dissipative resistor and filters are excluded based on the calibration. The bias drop $V$ also refers to the InAs-Al part,
excluding the film and filters. The error bars in Fig. 2f and Supplementary Information is estimated based on the
uncertainty of $R_{\text{series}}$ ($\sim 100$ $\Omega$) and the conductance variation of the ZBP near zero bias (within a window of 10 $\mu$V.)

\* equal contribution
† shangrn@baqis.ac.cn
‡ jhzhao@semi.ac.cn
§ hzquantum@mail.tsinghua.edu.cn

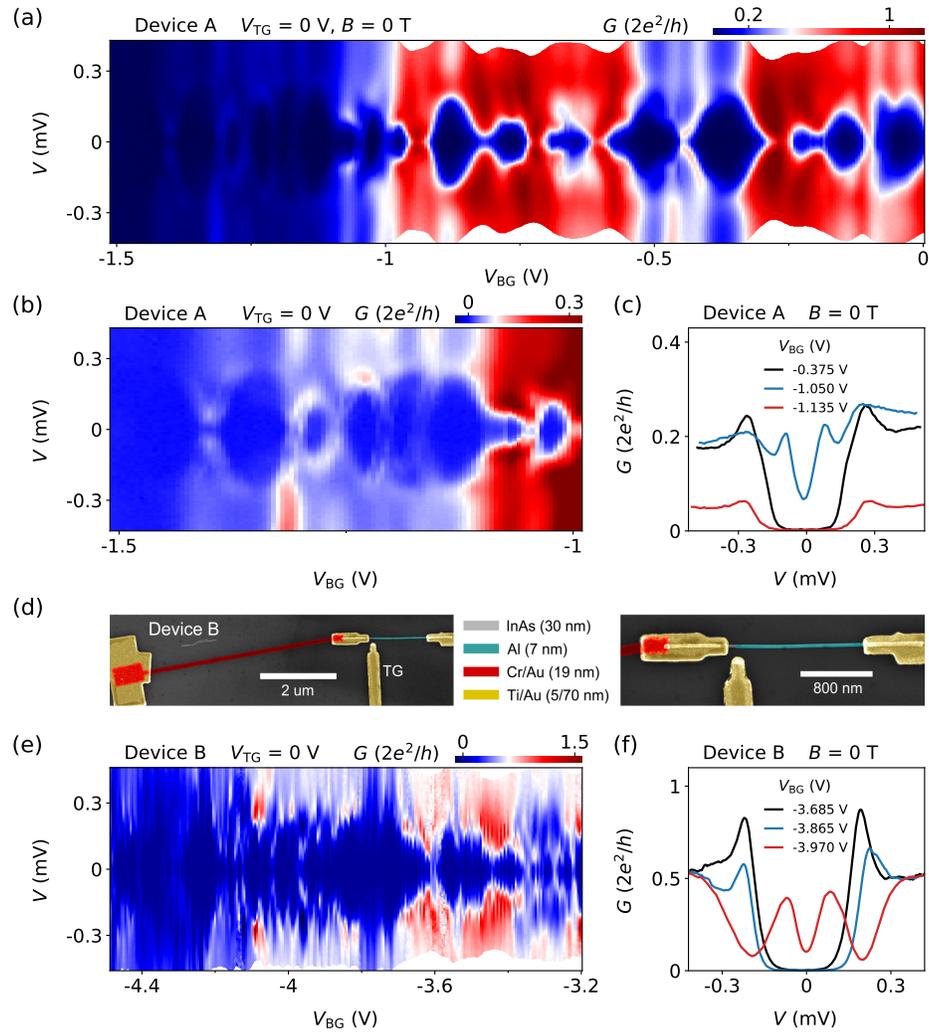

Fig. S 2: Pinch off calibration of Device A and B. (a) $G$ vs $V$ and $V_{BG}$ at $B = 0$ T for Device A. Disorder induced unintentional quantum dots and Coulomb blockades co-exist with the superconducting gap. (b) Zoom-in of the tunneling regime resolves sub-gap states. (c) Line cuts from (a-b). (d) False-color SEM of Device B. (e) 2D gate scan of Device B resolves more disorder induced quantum dots, suggesting a higher disorder level. (f) Line cuts from (e). $T \sim 24$ mK.

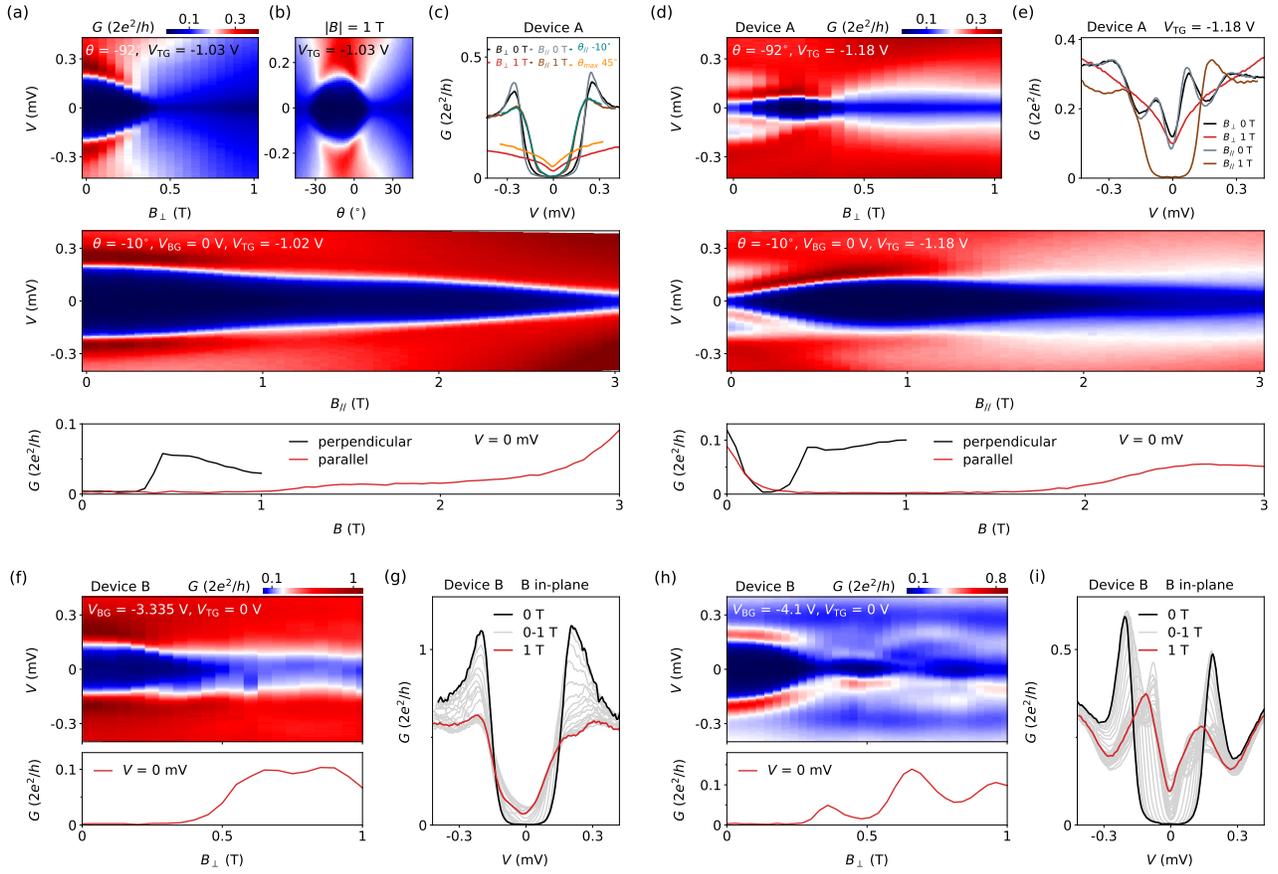

Fig. S 3: Signatures of environmental Coulomb blockade (ECB). (a-e) for Device A and (f-i) for Device B. (a) Upper, $B$ dependence of the superconducting gap with $B$ direction nearly perpendicular to the nanowire axis (by 8° misalignment). After the gap closing near $B = 0.5$ T, a conductance dip near zero bias is resolved, possibly due to ECB. Middle, $B$ direction aligned with the nanowire axis, i.e. $\theta = -10°$. The gap survives at high $B$. Lower, zero-bias line cuts from the upper and middle panels. (b) $G$ vs $V$ at a fixed $B$ magnitude ($|B| = 1$ T) with rotating field direction. The angle $\theta = -10°$ corresponds to the orientation of the nanowire axis. After ~30° rotation, the gap is closed and a ECB dip can be resolved. (c) Line cuts from (a-b) showing the gap and ECB dip. (d-e) Similar with (a) and (c) but at a different gate voltage setting where an ABS (the ground state being a doublet) can be resolved at low $B$. The co-existence of ABS does not change the gap-ECB transition. (f-i) $B$ dependence of the gap in Device B for two gate voltage settings, with $B$ perpendicular to the nanowire axis. Though a dip near zero bias could also be observed at high $B$, may be connected with ECB. Other features (e.g. unintentional dot states or Coulomb blockades) are also present, possibly due to a higher disorder level in this device. All the $B_\perp$ directions in this figure are in-plane.

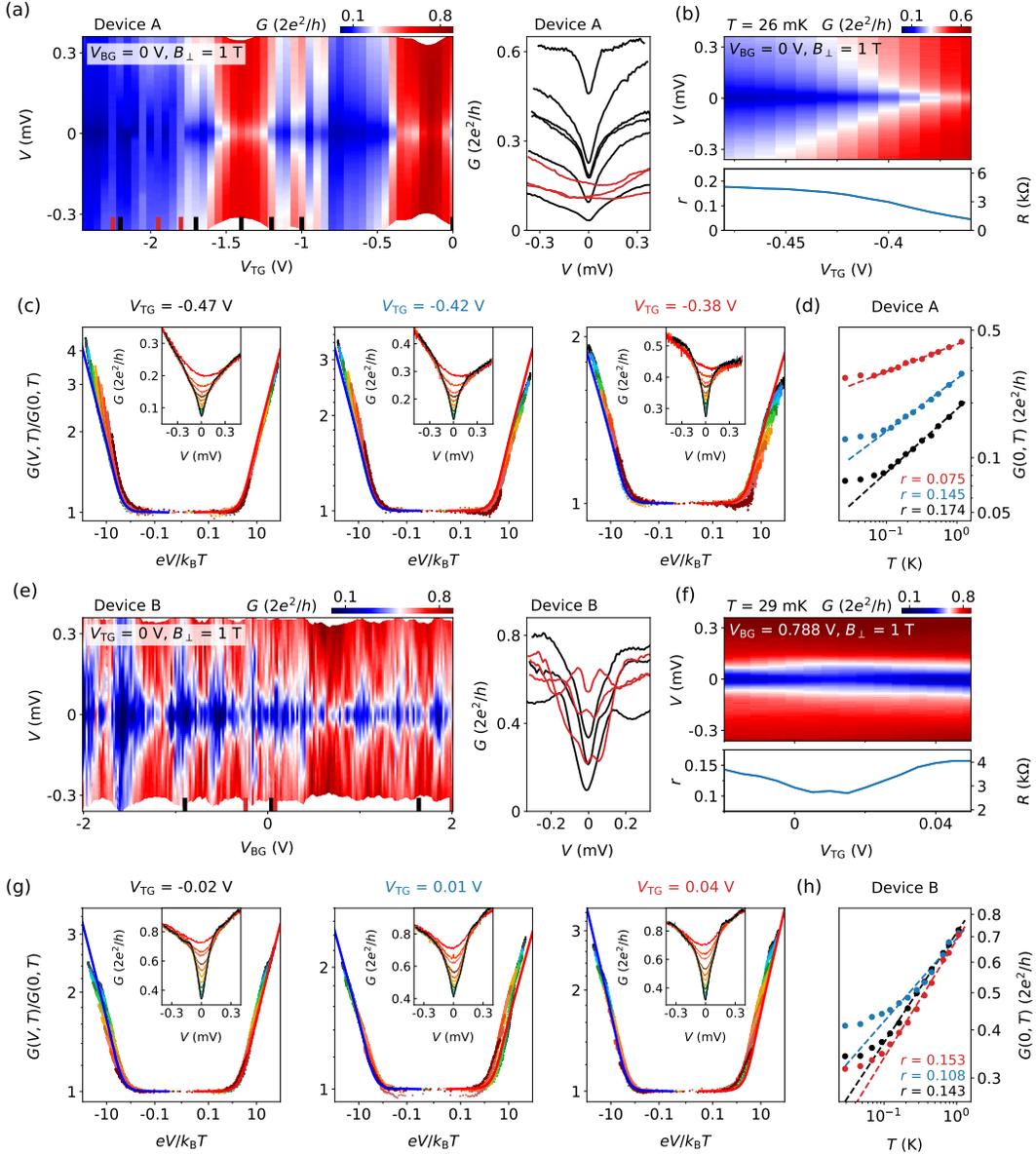

Fig. S 4: Power law of ECB in Device A and B. (a) $G$ vs $V$ and $V_{TG}$ for Device A at a perpendicular $B$ of 1 T (in-plane) to suppress its superconductivity. $T \sim 20$ mK. Most of the regions show a conductance suppression near zero-bias (see black line cuts on the right), while others do not (red line cuts) possibly due to the coexistence of states/features in the InAs-Al part. (b) Zoom-in on a 'clean' region from (a) for $T$ dependence. The measurement at base $T$ (upper panel) was performed for different $T$s (not shown) and the power law exponent $r$ was extracted for each gate voltage (lower panel), based on the $T$ dependence of zero-bias $G$. This exponent $r$ corresponds to an effective dissipation resistance: $R = r \times h/e^2$ (shown on the right axis). (c) $T$ dependence of three line cuts from (b). Insets show the line cuts (in linear scale) at different $T$s where higher $T$ lifts the ECB dip (the color-$T$ correspondence is similar to that in Fig. 3c). Main panels re-plot the insets using dimensionless units on a linear scale from -0.1 to 0.1 and a log scale outside this range for x-axis. $T$ in the re-scaled $x$-axis used the extracted electron $T$ (see panel d) for curves of $T < 100$ mK. All the line cuts roughly collapse onto a universal curve (blue and red lines), obtained based on $I(V, T) \propto V T^{2r} |\Gamma(r + 1 + ieV/2\pi k_B T)/\Gamma(1 + ieV/2\pi k_B T)|^2$ where $\Gamma$ is the Gamma function. (d) $T$ dependence of the zero-bias $G$ (dots) of (c). Dashed lines are the power law fits ($G \propto T^{2r}$) with exponents labeled. The saturation at low $T$ suggests a saturation of the device electron $T$ (lock-in excitation may also have an effect). (e-h) Power law analysis for Device B. $B_\perp$ is out-of-plane.

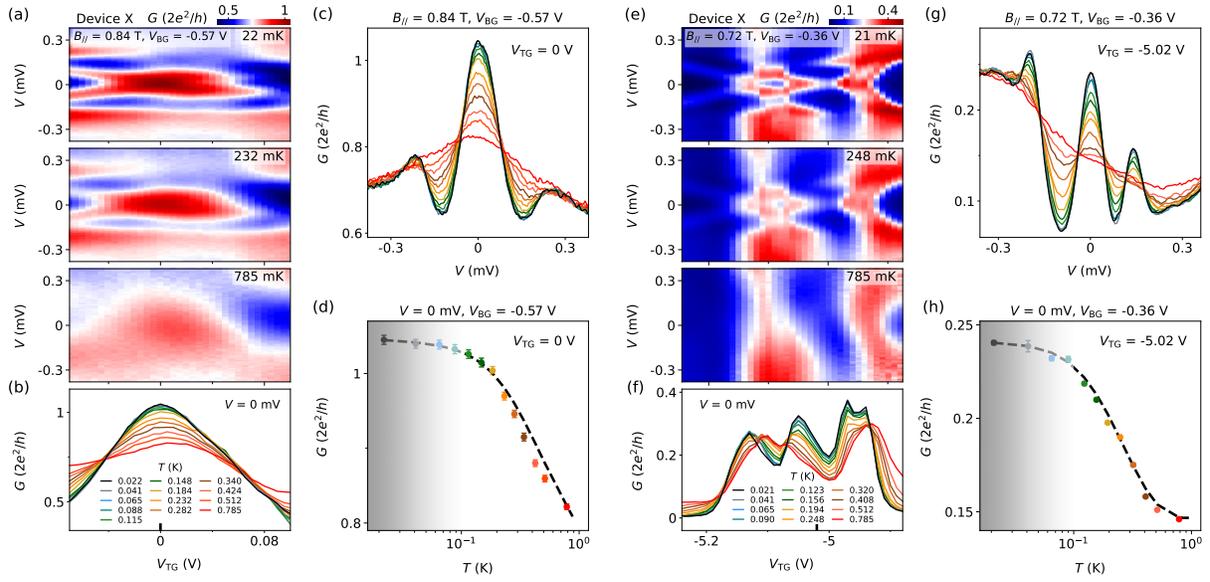

Fig. S 5: $T$ dependence of ZBPs in the control Device X, an InAs-Al nanowire without the dissipative resistor. (a) Gate scans of a zero-energy ABS at different $T$s (only three shown for clarity). (b) Zero-bias line cuts for all $T$s. (c) Line cuts of a ZBP for all $T$s. The peak height is near $2e^2/h$. (d) Zero-bias $G$ (dots) from (c) shows reasonably good match with thermal simulation (dashed line). (e-g) Another ZBP (with a smaller height) and its $T$ dependence at a different gate voltage, also showing good match with thermal simulation.

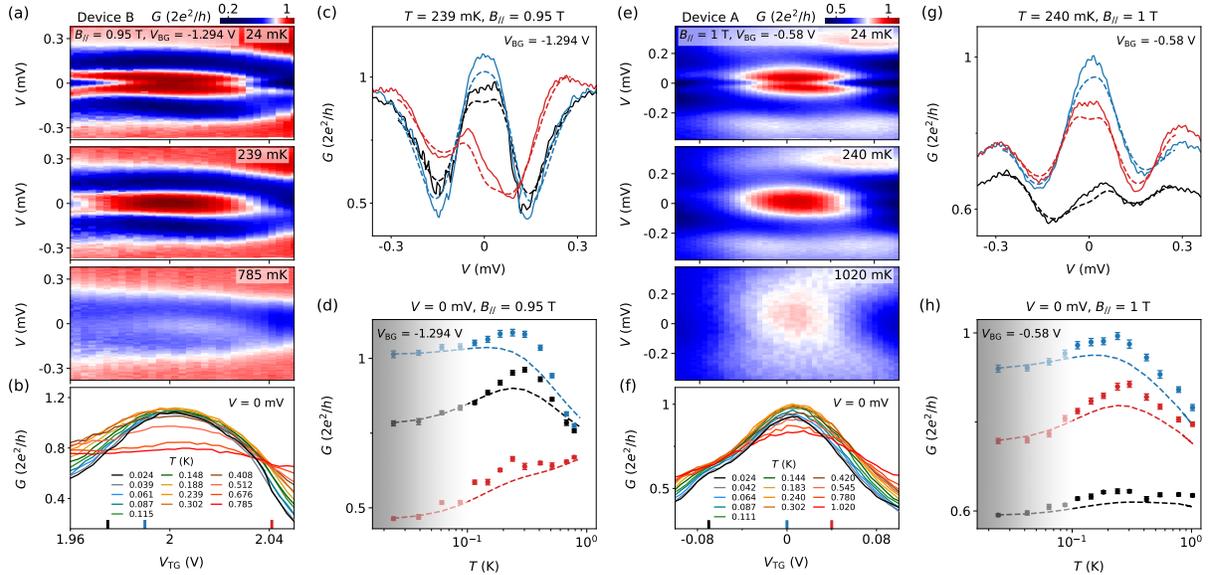

Fig. S 6: More $T$ dependence of the ABSs in Fig. 2. (a) Gate scans at different $T$s (only three shown for clarity) for Device B. (b) Zero-bias line cuts for all $T$s. (c) Line cuts of three zero- or near-zero-energy ABSs at $T = 239$ mK (colored lines) and the corresponding thermal simulations (dashed lines). The noticeable deviations indicate the effects of ECB. (d) Zero-bias $G$ of these line cuts (dots) and the thermal simulations (dashed lines) over the full $T$-range. The deviation is maximum for $T \sim 200 - 400$ mK. (e-h) same with (a-d) but for Device A.

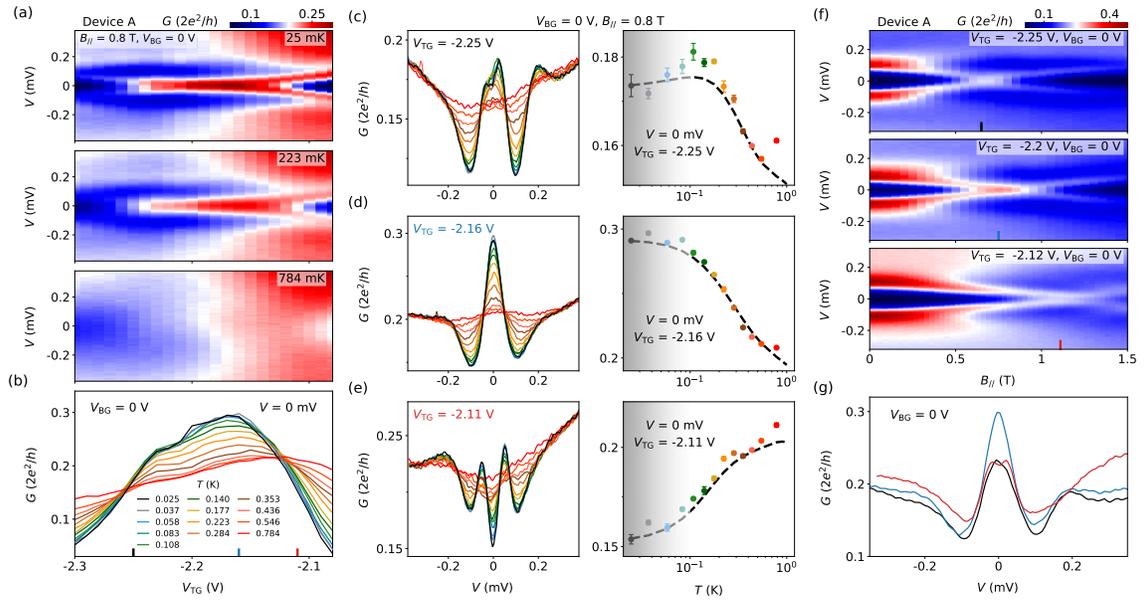

Fig. S 7: More on the ABS in Fig. 3. (a) Gate scans at three different $T$s. (b) Zero-bias line cuts for all $T$s. (c) $T$ dependence of three line cuts (left) and their zero-bias $G$ vs thermal simulation (right). Occasionally (top right panel), small deviations ($\sim 0.01 \times 2e^2/h$, slightly larger than the measurement noise) can be found possibly due to some instabilities at that particular gate voltage. (f) $B$ scans at three different gate settings with the zero-energy ABS line cuts shown in (g).

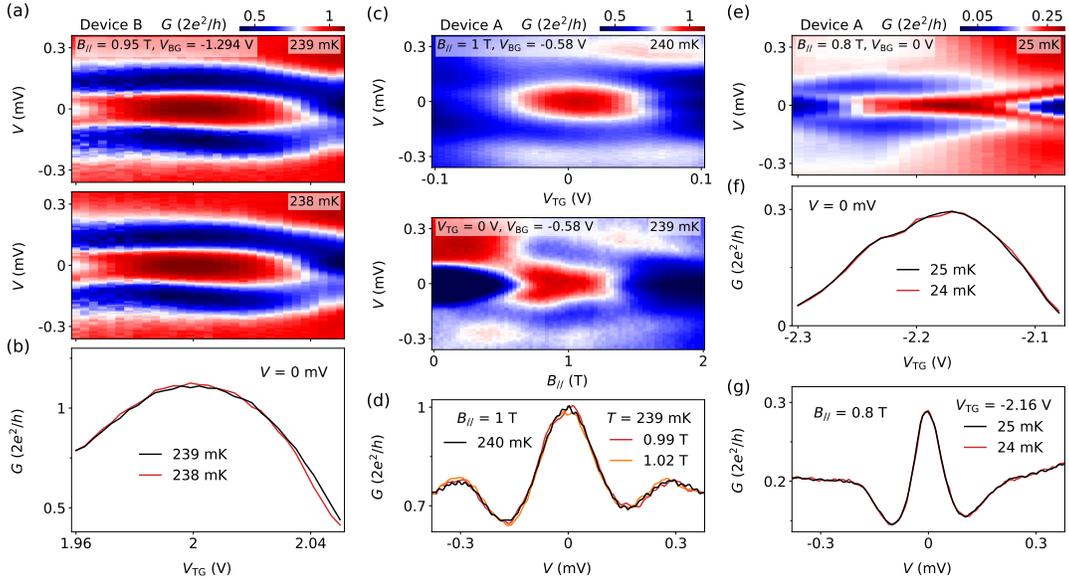

Fig. S 8: To rule out possible charge jumps, we have performed stability tests for the $T$ dependence measurement. (a) shows the test for Fig. 2c-f. The upper panel was measured at 239 mK in a $T$-sequence from low (24 mK) to high (785 mK). Afterwards, $T$ was then cooled down back to 238 mK and the measurement was repeated, shown in the lower panel. The good agreement with the re-measured data (line cuts shown in b) suggests no noticeable instabilities in between. (c) Upper (lower) panel is the same with the middle panel of Fig. 2i (Fig. 2g). Upper panel was measured in a $T$-sequence from 24 mK to 1.02 K, after which $T$ was cooled down back to 239 mK for the lower panel measurement. (d) Line cuts from the two panels of (c), corresponding to the same nominal parameter settings. The match suggests no noticeable instabilities in between. (e) Measurement performed right before Fig. 3b, with its zero-bias line cut shown in (f) (black), agreeing well with the red line (from Fig. 3b). (g) Black and red curves are the line cuts from (e) and Fig. 3a, sharing the same nominal parameter settings.

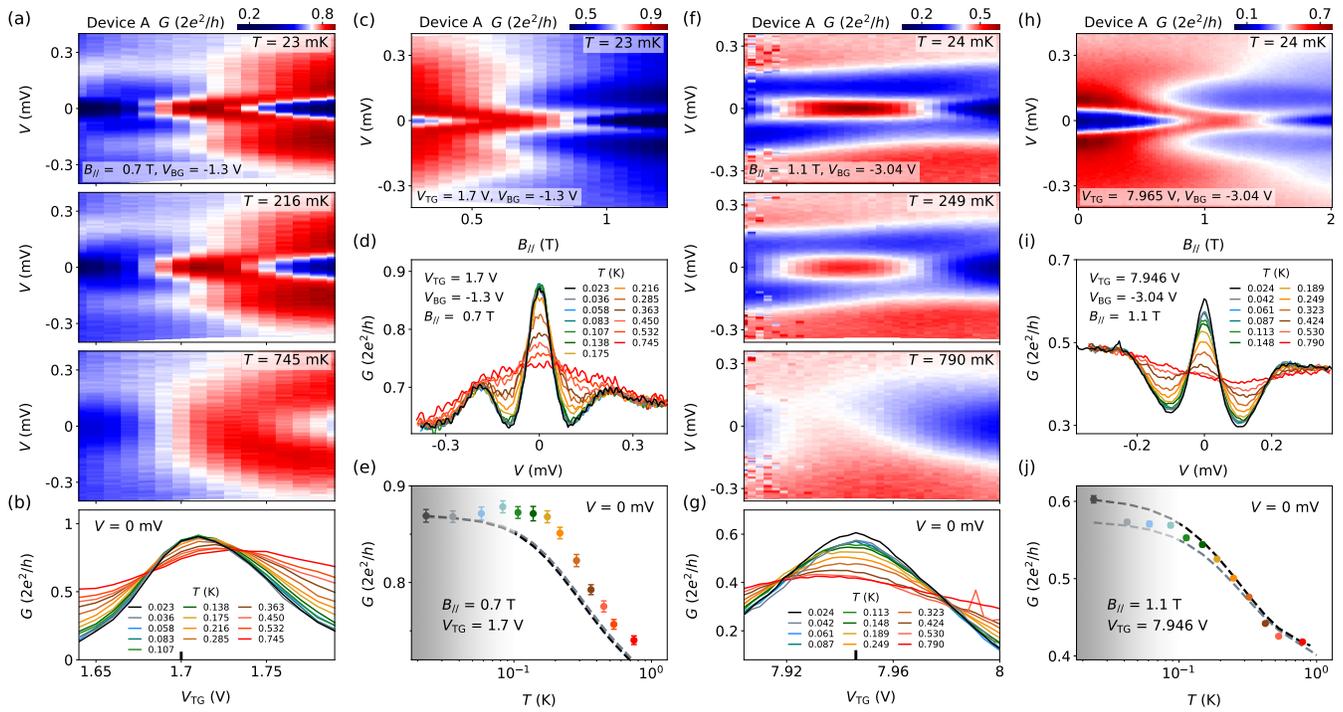

Fig. S 9: Two more zero-energy ABSs and $T$ dependence in Device A. (a) Gate scans at different $T$s (only three shown for clarity). (b) Zero-bias line cuts for all $T$s. (c) $B$ scan of the ZBP. (d) $T$ dependence of the ZBP line cut. (e) Zero-bias $G$ of (d) with thermal simulation (dashed line), showing noticeable deviations. (f-j) for another ABS. The two dashed lines in (j) are thermal simulations using the base $T$ and the second lowest $T$ as inputs to account for possible instabilities. The fitting in (j) shows less deviation with measurement compared to the case in (e). We note these two ABSs (together with Fig. S10-11) are the intermediate cases between Fig. 2 (large ZBPs) and Fig. 3 (small ZBPs), with a gradual transition from deviation to matching (between measurement and thermal simulation).

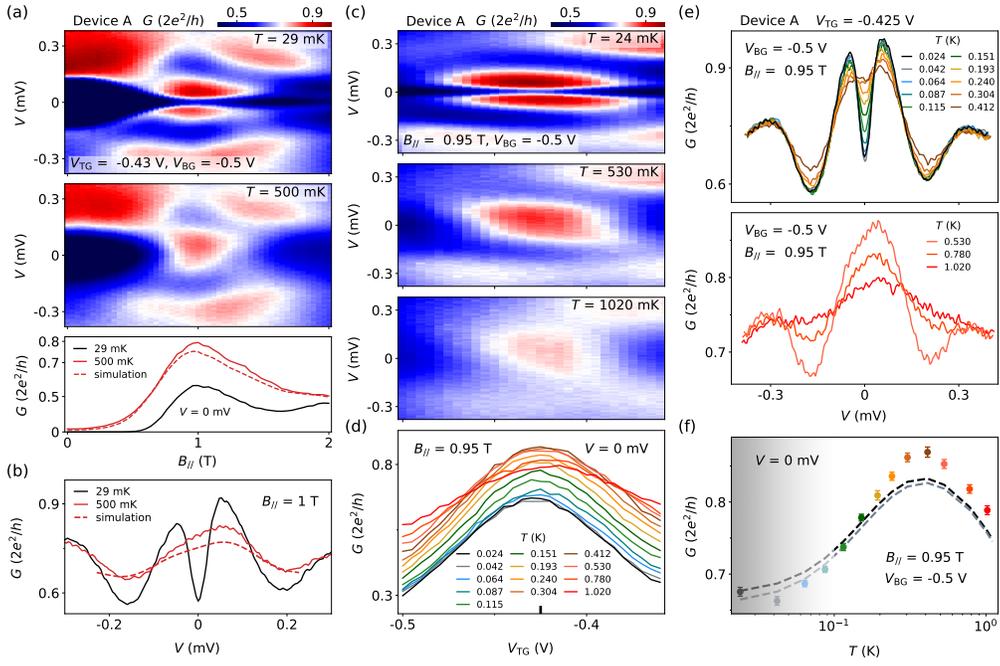

Fig. S 10: Another ABS in Device A. (a) $B$ scan of the ABS at 29 mK (upper) and 500 mK (middle). Zero-bias line cuts shown in the lower panel. The red dashed line is the thermal simulation for 500 mK. (b) Line cuts at 1 T. Thermal simulation for 500 mK shows noticeable deviations. (c) Gate scans at three different $T$s with zero-bias line cuts shown in (d). (e) $T$ dependence of a line cut from (c): the upper panel shows curves whose zero-bias $G$ increases with increasing $T$; the lower panel shows the opposite trend. (f) Zero-bias $G$ from (e) with thermal simulations (dashed lines).

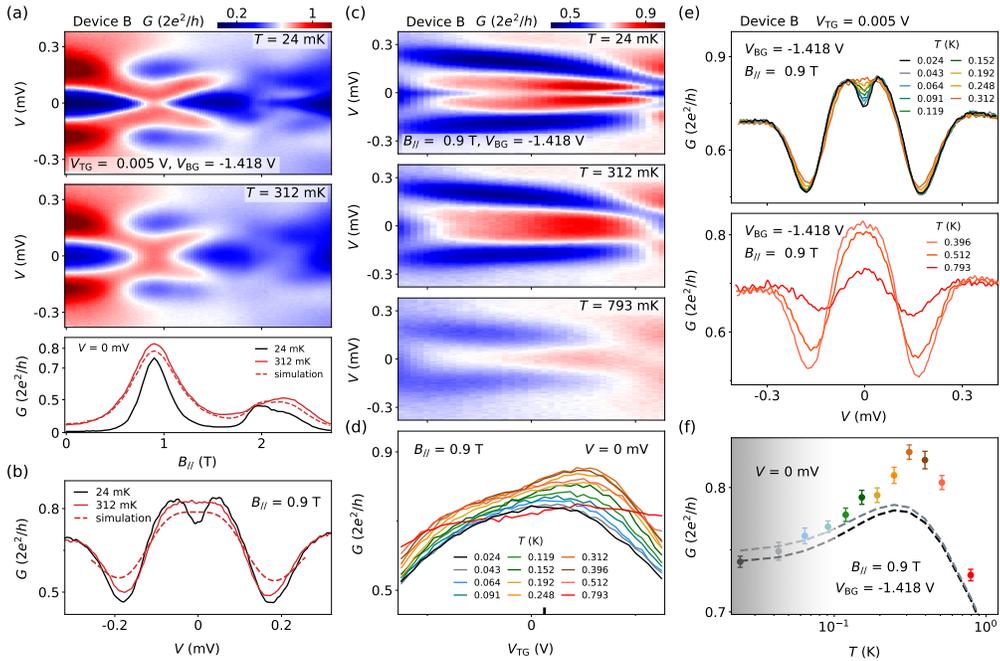

Fig. S 11: Another ABS in Device B. (a-f) are similar measurements with Fig. S10a-f.



\end{document}